\documentclass[%
reprint,
nofootinbib,
nobibnotes,
amsmath,amssymb,
aps,
prb,
floatfix,
]{revtex4-2}

\usepackage{graphicx}
\graphicspath{{figs/}}

\usepackage{bm}

\usepackage{newtxtext}
\usepackage[cmintegrals,cmbraces,smallerops,varg]{newtxmath}

\usepackage{ifpdf}
\usepackage{xcolor}
\definecolor{darkblue}{rgb}{0,0,.6}
\definecolor{darkgreen}{rgb}{0,0.5,0}

\ifpdf
  \usepackage{epstopdf}
  \usepackage[pdftex,unicode,pdfstartview={FitH},pdfborder={0 0 0}]{hyperref}
  \usepackage{hypcap}
\else
  \usepackage[hypertex]{hyperref}
\fi

\hypersetup{
  bookmarksnumbered = true,
  colorlinks = true, linkcolor = darkblue,
  citecolor = darkblue, filecolor = darkblue,
  menucolor = darkblue, urlcolor = darkblue
}

\usepackage{array}
\usepackage{tabularx}
\usepackage{makecell}
\usepackage{mathtools}
\usepackage{orcidlink}

\newcommand{\ii}{\mathrm{i}}
\DeclareMathOperator{\spn}{span}

\makeatletter
\newcommand{\llangle}{\langle\mkern-4mu\langle}
\newcommand{\rrangle}{\rangle\mkern-4mu\rangle}

\renewcommand{\Im}{\operatorname{Im}}
\newcommand{\Tr}{\operatorname{Tr}}

\DeclarePairedDelimiter\dket{|}{\rrangle}
\DeclarePairedDelimiter\dbra{\llangle}{|}

\DeclarePairedDelimiterX\dbraket[2]{\llangle}{\rrangle}{
  #1 \delimsize\vert #2
}

\DeclarePairedDelimiterX{\dmatel}[3]{\llangle}{\rrangle}{
  #1 \,\delimsize\vert\, #2 \,\delimsize\vert\, #3
}
\makeatother

\begin{document}

\title{Self-Consistent Spectral Quadrature Approach to\\Many-Body Green Functions}

\author{Stanislav Yu. Kruchinin \orcidlink{0000-0002-6495-3375}}

\email{stanislav.kruchinin@microsoft.com}
\affiliation{%
  Microsoft Austria\\
  Am Europlatz 3, 1120 Vienna, Austria
}%

\date{\today}

\begin{abstract}
We develop a self-consistent spectral quadrature (sc-SQ) framework for
the calculation of many-body Green functions.
The method approximates the K\"all\'en--Lehmann spectral measure by
Gauss--Christoffel (GC) quadrature, yielding a rational Green function
representation with guaranteed spectral positivity that exactly reproduces
the first $2N$ spectral moments at pole order $N$.
A key component is an SVD-based rank-selection criterion on the Hankel matrix,
which identifies the numerically resolvable pole rank $N^*$ from the
singular-value gap and acts as a precision-guided diagnostic of correlation
complexity.
The scheme is made self-consistent by requiring that the spectral function used to evaluate expectation values coincides with the spectral function generated by the quadrature reconstruction.
This defines a fixed-point hierarchy that connects systematically to established approximations, including Hartree--Fock and Hubbard-I, and incorporates non-perturbative features such as multi-peak spectral structure.
We benchmark the approach for the Anderson impurity model against
numerical renormalization group (NRG) results and apply it within dynamical
mean-field theory for the Hubbard model on the Bethe lattice.
The method captures the three-peak Anderson impurity spectrum and the
suppression of quasiparticle weight in the half-filled Hubbard model on
the Bethe lattice, including Mott-gap formation on the insulating branch
for $N\geqslant 5$, in qualitative agreement with NRG references.
\end{abstract}


\maketitle


\section{Introduction}
\label{sec:intro}

The calculation of many-body Green functions is central to
condensed matter physics and quantum chemistry, connecting
microscopic Hamiltonians to experimental observables such as spectral
functions, optical conductivities, and transport coefficients
\cite{Mahan2000, Mattuck1992}.
The central challenge is that the exact interacting Green function is
inaccessible in all but the simplest models, and every practical scheme
involves an approximation whose accuracy and physical validity must be
carefully assessed.
The most widely used framework is the $GW$ approximation and its
self-consistent extension (sc-GW) \cite{Aryasetiawan1998}, in which the
self-energy is approximated as $\Sigma = \ii GW$ with $W$ the dynamically
screened Coulomb interaction.
While $GW$ is highly successful for weakly correlated semiconductors and
simple metals, it suffers from well-documented structural failures in
strongly correlated materials: it violates spectral positivity in the
non-self-consistent $G_0W_0$ form, enforces self-consistency of a
diagrammatic functional truncated at second order in $W$, and does not
reliably capture Mott insulating behavior, a Kondo resonance, or
Hund multiplet splittings at practical levels of approximation~\cite{Aryasetiawan1998, Imada1998, Georges2013}.
These failures reflect the fact that strongly correlated phenomena are
non-perturbative in $W/U$ and therefore inaccessible to any method
organized as a power series in the screened Coulomb interaction.

The sc-SQ scheme proposed here operates on a fundamentally different
principle.
Rather than organizing the approximation as a power series in the interaction,
sc-SQ constrains the Green function to reproduce a finite set of exact spectral moments.
These moments encode interaction effects non-perturbatively through commutator algebra.
The resulting approximation enforces consistency between spectral reconstruction and moment evaluation,
providing a form of self-consistency distinct from the diagrammatic $\Phi$-derivable sc-GW scheme.
A detailed comparison of the two methods, supported by the benchmark
results of Sec.~\ref{sec:results}, is given in the Discussion
(Sec.~\ref{sec:discussion}).

To overcome the limitations of power-series expansions, rational
approximations, in particular Pad\'e approximants, have long been employed
as an effective tool for analytic continuation and series resummation
\cite{Baker1975, BakerGravesMorris}.
The $[N-1/N]$ Pad\'e approximant is uniquely suited for Green functions
because it naturally captures the required $1/z$ decay at high frequencies
and has been widely used for analytic continuation from the Matsubara axis
\cite{Vidberg1977, Beach2000, Schott2016}.
At the same time, the Nakajima--Mori--Zwanzig (NMZ) projection formalism \cite{Nakajima1958, Mori1965,
  Zwanzig1961} and the recursion method of Haydock, Heine, and Kelly
\cite{HaydockHeineKelly1972, Haydock1980} established that the exact Green
function in Liouville space can be written as a continued fraction whose
coefficients are determined by the Krylov dynamics of the Liouvillian
operator $\mathcal{L}[C] = [H, C]$.
Viswanath and M\"uller \cite{ViswanathMuller1994} provided the most
complete account of these connections in the condensed matter context,
demonstrating that the recursion method, the Lanczos algorithm, and the
theory of orthogonal polynomials are all facets of a single underlying
structure.

The equivalence of the Jacobi continued fraction $N$-th convergents to
$[N-1/N]$ Pad\'e approximants is a classical result in approximation
theory \cite{Gragg1972, BakerGravesMorris}. The guaranteed spectral
positivity of the resulting partial-fraction decomposition follows from
the theory of Gauss--Christoffel quadrature \cite{Brezinski1980, Beach2000},
which identifies the poles and residues as quadrature nodes and weights
for the spectral measure and thereby connects Haydock's recursion method
to the K\"all\'en--Lehmann spectral representation.
Lee's memory function formalism \cite{Lee1982} further showed that the
NMZ memory kernel plays the role of a spectral function of the
effective bath, a connection made rigorous in the quantum Brownian motion
framework of Grabert, Schramm, and Ingold \cite{Grabert1988} and equivalent
to a Caldeira--Leggett model \cite{CaldeiraLeggett1983} whose bath frequencies
and couplings are the Lanczos coefficients.

Despite the completeness of this formal picture, these connections have
not been assembled explicitly in the many-body literature, and a
self-consistent scheme that closes the loop between the rational
approximation and the moment input, thereby turning the hierarchy
into a practical many-body solver, has not previously been
formulated. The contributions of the present paper are threefold.

\textit{First}, we establish Gauss--Christoffel quadrature of the
K\"all\'en--Lehmann spectral measure as the organizing principle of the
sc-SQ framework. The $N$-th convergent of the Jacobi continued fraction
and the $[N-1/N]$ Pad\'e approximant emerge as equivalent representations
of the same GC construction rather than as independent approximation
schemes.

\textit{Second}, we introduce an SVD criterion for optimal pole-rank
selection.
A well-known practical obstacle to high-order recursion method is the
emergence of spurious pole-zero cancellations (Froissart doublets
\cite{Froissart1961}) caused by noise or truncation error in the perturbative
moments.
Existing approaches handle this by averaging over ensembles of Pad\'e
approximants constructed from randomly perturbed input \cite{Schott2016},
by \emph{ad hoc} pole-zero distance thresholds, or by data-driven
pole selection such as the AAA algorithm \cite{AAA2018}.
A complementary line of work represents spectral functions by a minimal
set of \emph{complex} poles optimized for a prescribed real-frequency
accuracy \cite{Zhang2025poles}. Because the poles are unconstrained in
the complex plane, spectral positivity is not guaranteed by construction.
Positivity-preserving analytic continuation via Nevanlinna functions
\cite{Fei2021} avoids spurious poles by construction but requires
Matsubara-axis input and a separate real-frequency extrapolation step.

We instead propose a criterion based on the singular value spectrum of the
Hankel moment matrix: the resolvable rank $N^*$ is identified as the number
of singular values exceeding a relative threshold tied to the numerical
precision of the moment input, below which singular values are treated as
noise rather than robust spectral content.
Beyond its role as a stabilizer, $N^*$ estimates the number of resolvable
many-body excitation channels and can jump sharply at correlation
crossovers, providing a cheap diagnostic of the nature of the correlated
state.

\textit{Third}, we develop the self-consistent
spectral quadrature (sc-SQ) scheme, in which the Gauss--Christoffel
reconstruction of the Liouvillian resolvent and the moment computation
are iterated to a self-consistent fixed point, and validate it against
the representative benchmarks.
The fixed-point hierarchy, comprising Hartree--Fock ($N=1$), Hubbard-I
($N=2$), the first central-resonance channel ($N=3$), and the onset of
multiplet resolution ($N=4$), places sc-SQ in a precise relationship with
established approximations, with each new pole activating a many-body
excitation channel absent at the previous level.

Complementary lines of work achieve compact rational representations
of Green functions using complex poles optimized for a prescribed
accuracy~\cite{Zhang2025poles, Leon2025} or compress imaginary-time
propagators via quantics tensor trains~\cite{Takahashi2025}.
These are data-driven fitting approaches, whereas sc-SQ reconstructs
a spectral function with guaranteed positivity, exact sum rules, and
all poles on the real axis.

\section{Spectral Quadrature Framework}
\label{sec:foundations}

We take the spectral measure of the retarded single-particle Green function,
rather than the Green function itself, as the primary object of approximation.
All structural guarantees --- positivity, sum-rule exactness, and the
minimality of the pole representation --- follow directly from this choice.

A causal Green function analytic in the upper half-plane admits the
K\"all\'en--Lehmann representation (Stieltjes transform)~\cite{Mahan2000}
\begin{equation}\label{eq:kallen_lehmann}
  G(z) = \int_{-\infty}^{\infty} \frac{A(\omega)}{z - \omega}\,d\omega,
\end{equation}
where $A(\omega) = -(1/\pi)\Im G(\omega+\ii 0^+) \geqslant 0$
is the spectral function satisfying $\int A(\omega)\,d\omega = 1$.

Expanding~\eqref{eq:kallen_lehmann} at large $z$ yields the moment sequence
\begin{equation*}
  \mu_n = \int \omega^n A(\omega)\,d\omega,
\end{equation*}
which encodes all information about $G(z)$ in a basis-independent form.

Recovering $G(z)$ from $\{\mu_n\}$ is therefore a moment problem for a
positive measure~\cite{Akhiezer1965, ShohatTamarkin1943}: find the positive
measure $A(\omega)\,d\omega$ whose Stieltjes transform is consistent with the
given data.

The unique $N$-point approximation to $A(\omega)\,d\omega$ that reproduces the
first $2N$ moments with strictly positive weights is provided by
Gauss--Christoffel (GC) quadrature~\cite{Gautschi2004, Brezinski1980}.
It replaces the continuous measure by the discrete one
\begin{equation}\label{eq:A_N_omega}
  A_{[N]}(\omega) = \sum_{i=1}^N w_i\,\delta(\omega - \epsilon_i) \geqslant 0,
\end{equation}
with nodes $\epsilon_i$ and weights $w_i > 0$ satisfying the $2N$ moment conditions
\begin{multline}\label{eq:sum_rules}
  \int_{-\infty}^{\infty} \omega^n\, A_{[N]}(\omega)\, d\omega
  = \sum_{i=1}^{N} w_i\,\epsilon_i^n = \mu_n,
\\
  n = 0, 1, \dots, 2N-1.
\end{multline}
The Stieltjes transform of~\eqref{eq:A_N_omega} gives the $N$-point rational
approximant
\begin{equation}\label{eq:pole_decomp}
  G_{[N]}(z) = \sum_{i=1}^{N} \frac{w_i}{z - \epsilon_i},
\end{equation}
which is by construction a Herglotz--Nevanlinna function (analytic
with non-positive imaginary part in the upper half-plane,
$\Im G_{[N]}(z) \leqslant 0$ for $\Im z > 0$).
Hereafter, we use the subscript index $[N]$ for the functions approximated by $N$ quadratures and their
equivalent forms.

Spectral positivity and exact sum rules~\eqref{eq:sum_rules} are thus structural
features of the GC approximation, not properties that need to be verified
separately.
As $N\to\infty$, the GC measure converges weakly to $A(\omega)\,d\omega$
whenever the moment problem is determinate~\cite{Akhiezer1965, ShohatTamarkin1943},
and $G_{[N]}(z) \to G(z)$.

The same approximant admits three equivalent representations that are
standard in the literature and useful in different computational contexts.
The nodes $\epsilon_i$ are the eigenvalues of the symmetric tridiagonal
(Jacobi) matrix~\cite{Gautschi2004, ViswanathMuller1994}
\begin{equation}\label{eq:tridiagonal_L}
  \mathbf{L}_N =
  \begin{pmatrix}
    a_0    & b_1    & 0      & \dots \\
    b_1    & a_1    & b_2    & \dots \\
    0      & b_2    & a_2    & \dots \\
    \vdots & \vdots & \vdots & \ddots
  \end{pmatrix},
\end{equation}
and the weights are $w_i = |(\mathbf{u}_i)_1|^2$, where $\mathbf{u}_i$ is the
normalized eigenvector of $\mathbf{L}_N$ corresponding to $\epsilon_i$.
The residues $w_i$ are strictly positive as reciprocals of sums of squares of
real orthogonal polynomials~\cite{Brezinski1980}, and are also given
by~\cite{BakerGravesMorris, ViswanathMuller1994}
\begin{equation}
  \label{eq:residues}
  w_i = \frac{P_{N-1}(\epsilon_i)}{Q'_N(\epsilon_i)},
\end{equation}
where $Q_N(z) = \det(z\mathbf{I} - \mathbf{L}_N)$ and $P_{N-1}(z)$ is the
associated numerator polynomial.

Equivalently, $G_{[N]}(z)$ is the $(0,0)$ matrix element of the resolvent
$(z - \mathbf{L}_N)^{-1}$, which generates the $N$-th convergent of the
Jacobi continued fraction (J-fraction)~\cite{BakerGravesMorris, Gragg1972,
ViswanathMuller1994}
\begin{equation}\label{eq:j_fraction}
  G_{[N]}(z) = \cfrac{1}{z - a_0 -
    \cfrac{b_1^2}{z - a_1 -
      \cfrac{b_2^2}{\ddots -
        \cfrac{b_{N-1}^2}{z - a_{N-1}}}}}.
\end{equation}

Finally, because $Q_N(z)$ reproduces the first $2N$ moments of $G(z)$, the
approximant coincides with the $[N-1/N]$ Pad\'e
approximant~\cite{Gragg1972, BakerGravesMorris}:
\begin{gather}\label{eq:pade_form}
  G_{[N]}(z) \equiv G_{[N-1/N]}(z) = \frac{P_{N-1}(z)}{Q_N(z)} = \frac{\sum_{i=0}^{N-1} p_i z^i}{\sum_{i=0}^N q_i z^i}
  ,\\ \nonumber
  G(z) - G_{[N]}(z) = O(z^{-(2N+1)}).
\end{gather}
The Pad\'e approximant is therefore not an independent construction but the
optimal rational representation induced by the GC quadrature of the spectral measure.
The Jacobi matrix~\eqref{eq:tridiagonal_L} and the continued
fraction~\eqref{eq:j_fraction} are likewise corollaries of the same structure.

For a many-body Hamiltonian $H$, the moments feeding the GC construction are
computed in Liouville space, where operators are treated as vectors with the
double-ket notation $\dket{C}$. The equilibrium inner product for fermionic
operators is
\begin{equation}
  \label{eq:inner_product}
  \dbraket{C}{D} = \langle \{C, D^\dagger\} \rangle = \Tr \left( \rho\, \{C, D^\dagger\} \right),
\end{equation}
where $\rho = e^{-\beta H}/Z$; for bosonic operators the anti-commutator is
replaced by the commutator throughout. The paper specializes to fermionic
$C = c_\sigma$, so $\mu_0 = \dbraket{C}{C} = 1$ by the canonical
anti-commutation relations. The Liouvillian superoperator
$\mathcal{L}\dket{C} \equiv \dket{[H,C]}$ is Hermitian with respect
to~\eqref{eq:inner_product}, and the retarded Green function takes the resolvent
form~\cite{ViswanathMuller1994}
\begin{equation}\label{eq:resolvent}
  G(z) = \dbra{C}(z - \mathcal{L})^{-1} \dket{C}, \qquad z = \omega + \ii 0^+.
\end{equation}
Expanding in powers of $1/z$ confirms that the moments are the Krylov inner
products
\begin{equation}\label{eq:moment_expansion}
  \mu_n = \dbra{C} \mathcal{L}^n \dket{C} = \langle \{[H,[\dots[H,C]\dots]], C^\dagger\} \rangle,
\end{equation}
computable as nested commutator expectation values in the interacting ground state~\cite{Mori1965, ViswanathMuller1994}.

There are four obvious routes to populate $\mathbf{L}_N$ and thereby define the
$N$-point GC rule.

(A)~\textit{Power moments}: compute $\mu_n = \dmatel{C}{\mathcal{L}^n}{C}$
by nested commutator algebra, form the $N \times N$ Hankel matrix
$\mathbf{M} = [\mu_{i+j}]_{i,j=0}^{N-1}$, and factor it as $\mathbf{M} = LDL^\top$;
the Lanczos recursion then yields $\mathbf{L}_N$ uniquely~\cite{ViswanathMuller1994,
Gragg1972}. Natural for local models where commutators close at finite order,
but the Hankel condition number grows as $O((N!)^2)$.

(B)~\textit{Direct Lanczos tridiagonalization}: restrict the dynamics to the
$N$-dimensional Krylov subspace
\begin{equation}
  \label{eq:krylov}
  \mathcal{K}_N = \spn\{\dket{C},\, \mathcal{L}\dket{C},\, \dots,\, \mathcal{L}^{N-1}\dket{C}\}
\end{equation}
and apply the Lanczos algorithm directly to $\dket{C}$, reading $\{a_i, b_i\}$
off the three-term recursion and bypassing $\mathbf{M}$ entirely. This avoids
the $O((N!)^2)$ condition-number growth and is the preferred route for $N\geqslant 4$.

(C)~\textit{Orthogonal polynomial expansion}: expand the spectral measure
in a family of polynomials $\{p_n\}$ orthogonal with respect to a reference
weight $\varrho(\omega)$ on $[-E_{\rm max}, E_{\rm max}]$, obtaining
expansion coefficients
$\alpha_n = \dmatel{C}{p_n(\mathcal{L}/E_{\rm max})}{C}$
that are bounded and numerically well-conditioned to high order.
The coefficients are converted to power moments via the standard
orthogonal-polynomial-to-monomial transform~\cite{ViswanathMuller1994},
after which the SVD criterion and Lanczos recursion proceed as in route~(A).
The polynomial family is a free parameter of the method, but Chebyshev
polynomials of the first kind $T_n$ ($\varrho \propto (1-x^2)^{-1/2}$)
are the recommended default because $|T_n(\cos\theta)| \leqslant 1$
uniformly and the expansion coefficients are computable via a fast
cosine transform~\cite{Weisse2006}.
Alternative families are preferable in specific contexts, but all of them 
produce the same $\{a_i, b_i\}$ when the input is exact,
since the three-term recurrence connecting any orthogonal family to the
Lanczos matrix is universal~\cite{Gautschi2004}.
This route is preferred for large sparse Hamiltonians and lattice models with
long-range hopping.

(D)~\textit{Green function sampling}: evaluate $G$ at a set of
probe points and fit a Stieltjes rational function to the samples.
Two natural sub-variants arise.
\textit{Frequency (Matsubara) sampling}: evaluate $G(z_k)$ at
probe frequencies on the imaginary axis or upper half-plane and solve
the resulting Cauchy-type interpolation problem~\cite{Fei2021, HuangAC2024};
natural for analytic continuation of Matsubara data at finite temperature.
\textit{Imaginary-time sampling}: evaluate $G(\tau_k)$
at discrete imaginary-time points, as produced directly by
finite-temperature quantum Monte Carlo (QMC), and fit via the Laplace-type
kernel $K(\tau,\omega) = e^{-\tau\omega}/(1+e^{-\beta\omega})$.
This extracts GC nodes directly from QMC data without a separate analytic continuation step~\cite{Huang2023}.
In both cases the SVD rank criterion is applied to the
corresponding kernel matrix (Cauchy or Laplace), and the fitting
problem reduces to the same GC quadrature structure.
All four routes produce the same $\{a_i, b_i\}$ when the input is exact,
they differ only in numerical conditioning and computational cost.

In route~(A), the Lanczos coefficients are
\begin{equation}
  \label{eq:lanczos_moments}
  a_i = \frac{\dmatel{v_i}{\mathcal{L}}{v_i}}{\dbraket{v_i}{v_i}}
, \quad
  b_i  = \sqrt{\frac{\dbraket{v_i}{v_i}}{\dbraket{v_{i-1}}{v_{i-1}}}},
\end{equation}
where $\dket{v_i} = (\mathcal{L} - a_{i-1})\dket{v_{i-1}} - b_{i-1}\dket{v_{i-2}}$
is the $i$-th Lanczos vector.

The diagonal elements $a_i$ are physical frequency shifts (Mori memory-function
frequencies~\cite{Nakajima1958, Mori1965, Lee1982}).
The off-diagonal $b_i$ measure inter-Krylov coupling and vanish when the recursion exhausts the physical
spectral weight, a fact underlying the SVD criterion of Sec.~\ref{sec:svd}.
For a two-body Hamiltonian, $\dket{v_n}$ is a $(2n+1)$-body operator, so the
$(0,0)$ element of $(z-\mathbf{L}_N)^{-1}$ yields the single-particle Green
function and each additional Lanczos step activates one higher many-body
process~\cite{Haydock1980, ViswanathMuller1994}.
The spectral reconstruction thus reduces to standard linear algebra:
Hankel factorization, Lanczos recursion, and symmetric tridiagonal eigendecomposition.

The proposed theoretical framework establishes three important connections that have previously been
described in different languages across the literature.

First, the equivalence between Haydock's recursion
method~\cite{HaydockHeineKelly1972, Haydock1980} and GC quadrature of the
K\"all\'en--Lehmann spectral measure has been previously noted in the
literature~\cite{Beach2000} but is stated here explicitly as the organizing
principle.

Second, the discrete spectral measure~\eqref{eq:A_N_omega} is formally
equivalent to a finite Caldeira--Leggett (CL) bath of $N$ harmonic
oscillators~\cite{CaldeiraLeggett1983}, with GC nodes $\epsilon_i$ as bath
frequencies and weights $w_i$ as coupling strengths. This is the many-body
analog of the chain mapping used in tensor-network methods for open quantum
systems~\cite{Chin2010, deVega2015}, where a continuous bath is likewise mapped
to a discrete harmonic chain by Lanczos orthogonalization.

Third, the NMZ projection~\cite{Nakajima1958, Mori1965, Zwanzig1961} 
yielding the continued-fraction representation of the Green function
produces a generalized Langevin equation for $C$ whose memory kernel $K(t-t')$
is the bath correlation function of this CL model, with parameters determined
by $\{a_i, b_i\}$ via the tridiagonal eigendecomposition~\cite{Grabert1988}.

The exact guarantees of the construction should be distinguished from the
approximations used to make it a practical solver. For any supplied
positive moment sequence, the GC step produces real poles, positive weights,
and exact reproduction of the first $2N$ moments of that supplied sequence.
The quality of those moments, the closure used for higher-order correlators,
the finite-precision SVD truncation, and any subsequent broadening or
perturbative linewidth correction are separate numerical or physical
approximations. The benchmarks below therefore compare the full practical
scheme, while the positivity and finite-moment guarantees apply strictly to
the bare GC/sc-SQ Green function before optional post-processing.

\section{SVD Rank-Selection Criterion}
\label{sec:svd}

In practice, the moments $\mu_n$ are not known exactly but are computed
perturbatively or via equations of motion, and their numerical errors amplify
rapidly with $n$. This leads to near-singular Hankel matrices $\mathbf{M}$
and to the appearance of Froissart doublets \cite{Froissart1961}, which are
spurious pole-zero cancellations that carry no physical spectral weight but
can distort the reconstructed spectral function
significantly. Existing stabilization strategies, such as averaging over an
ensemble of Pad\'e approximants constructed from randomly perturbed input
\cite{Schott2016}, are effective but require tuning an ensemble size and a
perturbation amplitude.

Regardless of the initialization route, the $N\times N$ matrix encoding
the quadrature problem --- the Hankel matrix $\mathbf{M}$ (routes~A and~C), the Gram matrix of Krylov vectors (direct Lanczos),
or the Cauchy/Laplace kernel matrix (Green function sampling) --- has a
singular value spectrum that reflects the resolvable numerical rank of the
spectral measure for the supplied data.
We state the criterion for the Hankel case. The extension to other input
formats is immediate.

We propose a more direct criterion based on the singular value decomposition
(SVD) of $\mathbf M$. Let $\mathbf M = U\Sigma V^\top$ with singular values
$\sigma_1 \geqslant \sigma_2 \geqslant \cdots \geqslant \sigma_N \geq 0$.
The numerically resolvable rank of the approximant is defined as
\begin{equation}\label{eq:svd_criterion}
  N^* = \max\!\left\{ n : \sigma_n / \sigma_1 > \tau \right\},
\end{equation}
where $\tau$ is a dimensionless threshold chosen from the accumulated floating-point
error in the moment computation and from the conditioning of the input matrix.

The recommended approximation is then $G_{[N^*]}(z)$, i.e., the
series~\eqref{eq:pole_decomp} truncated after $N^*$ steps.
We note that high-frequency stabilization of Green functions
and self-energies in DMFT remains an open problem~\cite{LaBollita2025}.
The present SVD criterion offers a complementary approach that links
stabilization directly to the preservation of exact spectral sum rules
rather than to constrained optimization of a few asymptotic coefficients.

A small singular value $\sigma_n \ll \sigma_1$ signals, within the accuracy of
the supplied moments, that
$b_{n-1}$ is numerically zero: the Krylov dynamics has terminated,
and retaining poles beyond $N^*$ introduces Froissart doublets.
Truncating at $N^*$ ensures real, positive-weighted poles and that
sum rules~\eqref{eq:sum_rules} hold to within the precision of the
input moments.

The SVD criterion~\eqref{eq:svd_criterion} requires one numerical threshold
$\tau$ whose natural scale is set by moment precision, unlike ensemble
averaging (two hyperparameters) or pole-zero distance thresholds
(a system-dependent length scale).
It also differs from the AAA algorithm~\cite{AAA2018}, which selects
poles by greedy residual minimization of a known function. AAA is
data-driven, whereas the SVD criterion preserves the maximum number
of exact spectral sum rules from moment input alone.
The condition number of the Hankel matrix $\mathbf{M}$ grows as
$O((N!)^2)$ with $N$~\cite{GolubVanLoan2013}, so the SVD threshold
$\tau$ must be set well above floating-point round-off.
The value $\tau = 10^{-8}$ used throughout is conservative relative to both
the machine-precision floor and the smallest retained singular value in the
benchmarks.

\section{Self-Consistent Scheme}
\label{sec:scpade}

\subsection{The self-consistency condition}
\label{sec:sc_condition}

The input parameters of the quadrature rule (whether spectral moments
$\mu_n = \dmatel{C}{\mathcal{L}^n}{C}$ or Lanczos coefficients $\{a_i, b_i\}$
obtained by direct tridiagonalization) are not independent of the Green
function: for $n \geqslant 2$ they involve expectation values of products
of operators whose exact values depend on the interacting ground state.
In the one-shot scheme of the preceding sections these expectation values
are supplied from an external source (a non-interacting reference state
or a low-order perturbative calculation), and the quadrature reconstruction
is performed once. The one-shot approach is accurate when the reference state
is close to the true ground state, but it fails in the strongly correlated
regime precisely where the hierarchy is most needed.

The self-consistency condition~\eqref{eq:fixed_point_map} is stated below
in terms of spectral moments because they provide the simplest closed-form
expression of the fixed-point requirement. When direct Lanczos is used as
the initialization route, the analogous condition is that the Lanczos
coefficients $\{a_i^{(\ell)}, b_i^{(\ell)}\}$ computed at iteration $\ell$
from the current spectral function equal those produced by applying the
Lanczos recursion to the updated Hamiltonian with the new expectation
values, a formally identical but numerically better-conditioned statement
of the same fixed point.

The sc-SQ method closes this gap by requiring the spectral function
used in the moment closure to be the same as the spectral function
produced by the quadrature reconstruction. Single-particle expectation
values are obtained directly from the one-particle spectral function.
Higher correlators entering $\mu_n$, such as double occupancy or
spin/charge correlators, require either EOM identities, two-particle
information, or an explicit closure. We write the resulting closure in the
generic form
\begin{equation}
  \label{eq:sc_condition}
  \langle B \rangle = \int_{-\infty}^{0} K_B(\omega)\, A(\omega)\, d\omega,
\end{equation}
where $K_B(\omega)$ is the kernel implied by the chosen EOM relation or
closure for the observable $B$, and $A(\omega)$
is the interacting spectral function used in that closure. For example, the occupancy
$\langle n_\sigma \rangle = \int_{-\infty}^{0} A(\omega)\,d\omega$
corresponds to $K_{n_\sigma}(\omega) = 1$, while the double occupancy
$\langle n_\uparrow n_\downarrow \rangle$ requires a two-particle kernel
or an EOM-based approximation, rather than following from the one-particle
spectrum alone \cite{Mahan2000, Mattuck1992}.

Substituting the pole decomposition \eqref{eq:pole_decomp} into
\eqref{eq:sc_condition} converts all integrals into finite sums:
\begin{equation}\label{eq:sc_discrete}
  \langle B \rangle^{(\ell)} = \sum_{i:\,\epsilon_i^{(\ell)} < 0} w_i^{(\ell)}\,K_B\! \left(\epsilon_i^{(\ell)}\right),
\end{equation}
where the superscript $(\ell)$ labels the iteration. The moments
$\mu_n^{(\ell + 1)}$ are then recomputed from the updated expectation values,
a new Hankel matrix is formed, the SVD criterion \eqref{eq:svd_criterion}
is applied to determine $N^{*(\ell + 1)}$, and a new set of poles and weights
$\{\epsilon_i^{(\ell + 1)}, w_i^{(\ell + 1)}\}$ is obtained. The cycle is repeated
until convergence.

\subsection{Iterative algorithm}
\label{sec:algorithm}

The sc-SQ iteration at order $N$ proceeds as follows.

\textit{Initialization.} Compute the moments $\mu_0, \dots, \mu_{2N-1}$
using expectation values from a non-interacting or Hartree-Fock reference
state. This defines the starting approximant $G^{(0)}_{[N]}$.

\textit{Iteration.} Given the poles $\{\epsilon_i^{(\ell)}\}$ and weights $\{w_i^{(\ell)}\}$ from step $\ell$:
\begin{enumerate}
  \item Evaluate all required expectation values via Eq.~\eqref{eq:sc_discrete}.
  \item Recompute the input parameters of the quadrature rule from the
  updated expectation values via the initialization route of
  Sec.~\ref{sec:foundations} (power moments, direct Lanczos, orthogonal
  polynomial expansion, or Green function sampling), and form $\mathbf{L}_N^{(\ell+1)}$.
  \item Apply the SVD criterion~\eqref{eq:svd_criterion} to the appropriate
  input matrix (Hankel, Gram, Cauchy, or Laplace, depending on the
  initialization route) to determine $N^{*(\ell+1)}$.
  \item Diagonalize $\mathbf{L}_{N^{*(\ell+1)}}^{(\ell+1)}$ to obtain
  $\{\epsilon_i^{(\ell+1)}, w_i^{(\ell+1)}\}$.
\end{enumerate}

\textit{Convergence criterion.} The computational process is considered converged when
\begin{equation}\label{eq:convergence}
  \max_n \frac{\left| \mu_n^{(\ell + 1)} - \mu_n^{(\ell)} \right|}{\left| \mu_n^{(\ell)} \right|} < \delta,
\end{equation}
with $\delta = 10^{-8}$ in practice.

The zeroth moment $\mu_0 = 1$ is exact by the anticommutation relations and never changes.
The first moment $\mu_1 = \varepsilon_d + U\langle n_{\bar\sigma}\rangle$ depends on the
occupancy and is updated at each iteration, but it is a simple
single-operator expectation value rather than a connected correlator.

The self-consistency is therefore driven primarily by the second and
higher moments, which contain the non-trivial many-body correlations.
The per-iteration cost is $O(N^3)$ for the Lanczos
diagonalization plus the cost of evaluating the expectation values in
step 1, which is $O(1)$ for local models and $O(N_k)$ for
lattice models with $N_k$ quasimomentum points.

Throughout the iteration we track three diagnostics: $N^{*(\ell)}$,
$\langle n_\sigma\rangle^{(\ell)} = \sum_{i:\epsilon_i^{(\ell)}<0} w_i^{(\ell)}$,
and $\langle n_\uparrow n_\downarrow\rangle^{(\ell)}$.
Stabilization of $N^{*(\ell)}$ is a necessary condition for a genuine
fixed point of the self-consistency map~\eqref{eq:fixed_point_map}
(Sec.~\ref{sec:thermodynamic}): oscillating rank signals cycling
between approximants rather than convergence.
The occupancy and double occupancy serve as thermodynamic sentinels
and are used to verify the consistency condition in Sec.~\ref{sec:thermodynamic}.

\subsection{Fixed-point analysis and connection to known approximations}
\label{sec:fixed_point}


The sc-SQ approximation is controlled by the number of preserved moments,
not by the strength of the interaction: increasing $N$ monotonically
tightens the constraints on the spectral measure.
For the one-shot hierarchy, the convergence $G_{[N]}\to G$ as $N\to\infty$
follows from the theory of Gauss--Christoffel quadrature whenever the
spectral moment problem is determinate and the moments are noise-free.
For the self-consistent fixed-point sequence $\{G^*_{[N]}\}_{N=1}^\infty$
this convergence is a physically motivated assumption rather than a
proved theorem: the fixed-point map $\mathcal{F}$ itself changes with $N$,
and no general monotone convergence result is available.
In practice the sequence converges rapidly for the models studied here,
and the benchmark comparisons with NRG provide the empirical validation.
This non-perturbative character allows the method to represent
Mott-gap formation, central Kondo-resonance precursors, and multiplet
splittings through moment constraints rather than through any finite-order
diagrammatic expansion in the screened Coulomb interaction.

The structure of the sc-SQ scheme places it in a precise
relationship with several well-established approximations.
We now make this explicit for each level of the hierarchy.

\textit{$N=1$: Hartree--Fock approximation.}
The 1-point GC rule is fixed by the two lowest moments, $\mu_0 = 1$ and
$\mu_1 = \langle \{[H, c_\sigma], c_\sigma^\dagger\} \rangle$.
The sum-rule conditions~\eqref{eq:sum_rules} have the unique solution
$\epsilon_1 = \mu_1$ with weight $w_1 = 1$, representing one excitation
channel: the bare electron propagating in the mean field of all others.
The self-consistency condition requires $\epsilon_1
= \varepsilon_d + U\langle n_{\bar\sigma}\rangle$, where
$\langle n_{\bar\sigma}\rangle$ is evaluated from the single-node
spectral measure via Eq.~\eqref{eq:sc_discrete}. This is precisely the
Hartree--Fock equation: sc-SQ at $N=1$ is therefore identical to
self-consistent Hartree--Fock, providing a clean base case. A
non-interacting system is described exactly at this level and the GC
rule terminates at $N^*=1$.

\textit{$N=2$: Hubbard-I approximation and the two-channel picture.}
The 2-point GC rule is determined by the four moments $\mu_0,\dots,\mu_3$.
The second node activates the second excitation channel: the added electron
encounters a site that is either already occupied (energy cost $\sim U$) or
empty (energy $\sim \varepsilon_d$), corresponding to the upper and lower
Hubbard bands. For the Anderson impurity at particle-hole symmetry
($\varepsilon_d = -U/2$, $\langle n_\sigma \rangle = 1/2$), $\mu_3$
involves the double occupancy $\langle n_\uparrow n_\downarrow\rangle$.
With the Hubbard-I factorization $\langle n_\uparrow n_\downarrow\rangle
= \langle n_\uparrow \rangle \langle n_\downarrow \rangle = 1/4$,
the $2N=4$ sum-rule conditions~\eqref{eq:sum_rules} determine the two
nodes as
\begin{equation}\label{eq:hubI_poles}
  \epsilon_\pm = \frac{\mu_1}{\mu_0} \pm
  \sqrt{\frac{\mu_2}{\mu_0} - \left(\frac{\mu_1}{\mu_0}\right)^2},
\end{equation}
which reduce to $\epsilon_\pm = \pm U/2$, the exact Hubbard-I band
positions, with equal weights $w_\pm = 1/2$, in the atomic limit
  $V_k \to 0$. With finite hybridization the nodes shift to
  $\pm\sqrt{U^2/4 + \sum_k |V_k|^2}$, consistent with spectral weight
  transfer to the Hubbard bands. The sc-SQ fixed point at $N=2$ is
therefore the Hubbard-I approximation.

\textit{$N=3$: Central resonance channel and dynamical screening.}
The third quadrature node activates a qualitatively new channel: the
added electron can resonantly exchange spin with the conduction bath,
the process underlying Kondo screening. This process is \textit{dynamical}, not captured
by any static mean-field or Hubbard-I factorization, and it is first
encoded in $\mu_4$, which contains the connected four-operator correlator
$\langle n_\uparrow n_\downarrow c^\dagger_\sigma c_\sigma \rangle_c$
measuring quantum fluctuations of the double occupancy. At $N=3$
the sc-SQ fixed point evaluates this correlator self-consistently from the
three-node GC measure, without the Hubbard-I factorization, and thereby
captures exchange and correlation corrections to the node positions and
weights that lie entirely beyond the Hubbard-I level.
The physical consequence is the appearance of a central node between the
two Hubbard-band nodes. In the Anderson model this central node carries
the precursor of the Abrikosov--Suhl resonance: its position is pinned
to the Fermi level by particle-hole symmetry, while its weight provides
a finite-rank measure of the low-energy spectral weight. Quantitative
extraction of the exponentially small Kondo scale over many decades
requires higher ranks than those benchmarked here.
For single-orbital models at moderate
coupling, $N=3$ captures the essential three-peak structure of the
correlated metal and is typically sufficient.

\textit{$N=4$: Inelastic scattering and the first satellite.}
The fourth quadrature node activates the first inelastic scattering
channel: a shake-up satellite at energy offset $\sim U$ from the
quasiparticle peak, with spectral weight growing with coupling strength.
The new information enters through $\mu_6$ and $\mu_7$, the first moments
to contain six-operator correlators. In multi-orbital models with Hund
coupling $J$, the fourth node also begins to resolve the $S=0$ and $S=1$
multiplet configurations degenerate at $N=3$, a precursor of
orbital-selective Mott physics~\cite{Georges2013}. More generally,
$N^*=1$ for non-interacting systems, $N^*=1$--$2$ for Fermi liquids,
and $N^*$ grows with active orbital channels in Hund metals, so $N^*$
counts resolvable excitation channels.

\subsection{Spectral function reconstruction}
\label{sec:spectral_reconstruction}

The converged sc-SQ Green function $G_{[N^*]}(z)$ is a rational function
with $N^*$ simple poles on the real axis, so the corresponding spectral
function
\begin{equation}
  A_{[N^*]}(\omega) = \sum_{i=1}^{N^*} w_i\,\delta(\omega - \epsilon_i)
\end{equation}
is a discrete measure.
This is an intermediate quantity whose physical content resides entirely
in the pole positions $\{\epsilon_i\}$ and weights $\{w_i\}$: as argued
in Sec.~\ref{sec:fixed_point}, these are the Gauss--Christoffel
quadrature nodes and weights for the true spectral measure, and their
convergence toward the true excitation energies and quasiparticle
residues as $N\to\infty$ is a well-defined statement independent of any
broadening.

Physically realistic lineshapes are obtained from $A_{[N^*]}(\omega)$ by one of the following
procedures, depending on the context.
Of these, SEC is used for the Anderson impurity benchmarks of
Sec.~\ref{sec:bench_anderson}, with an optional perturbative linewidth
correction described there. For the Bethe lattice DMFT benchmarks of
Sec.~\ref{sec:bench_hubbard}, the finite-rank poles are broadened
with Lorentzians for visualization, while pole positions and weights remain
the primary sc-SQ outputs.

\textit{Self-energy continuation (SEC).}
Within DMFT, the natural route is to extract the
self-energy from the converged pole decomposition,
\begin{equation}
  \label{eq:selfenergy}
  \Sigma_{[N^*]}(z) = \mathcal{G}_0^{-1}(z) - G_{[N^*]}^{-1}(z),
\end{equation}
and evaluate the physical Green function on the real axis as
\begin{equation}\label{eq:Gphys}
  G_{\text{phys}}(\omega + \ii\eta)
  = \bigl[\omega + \ii\eta - \varepsilon_d
  - \Delta(\omega+\ii\eta) - \Sigma_{[N^*]}(\omega+\ii\eta)\bigr]^{-1},
\end{equation}
where $\Delta(z)$ is the bath hybridization function.

Because $\Delta(\omega + \ii\eta)$ has a continuous imaginary part for any
$\eta > 0$, the resulting spectral function
\begin{equation}\label{eq:A_SEC}
  A_{\text{SEC}}(\omega, \eta) = -\frac{1}{\pi}\Im G_{\text{phys}}(\omega+\ii\eta)
\end{equation}
is automatically smooth with physically meaningful quasiparticle
linewidths set by the bath, with no artificial broadening introduced.
This is the procedure used for the Anderson impurity benchmarks of
Sec.~\ref{sec:bench_anderson}.

\textit{Lorentzian broadening (LB).}
For impurity benchmarks outside DMFT, where no bath hybridization
provides intrinsic broadening, the discrete poles are convolved with
a Lorentzian of half-width $\eta$:
\begin{equation}\label{eq:broadening}
  A_{\text{LB}}(\omega, \eta) = \frac{1}{\pi}\sum_{i=1}^{N^*} \frac{w_i\,\eta}{(\omega-\epsilon_i)^2 + \eta^2}.
\end{equation}

Here $\eta$ is a phenomenological parameter, not a physical one.
All qualitative features (pole count, peak positions, relative weights) are insensitive
to $\eta$ in the range $0.01$--$0.05\,\Delta$.
The physically meaningful quantities reported for quantitative comparison
with NRG are always the poles $\{\epsilon_i\}$ and weights $\{w_i\}$
directly, not the broadened lineshape.

\textit{Barycentric rational interpolation (BRI).}
Because $\{\epsilon_i, w_i\}$ are Gauss--Christoffel nodes and weights, they
implicitly define a unique barycentric rational interpolant~\cite{Berrut2004, Wang2014}
that gives a smooth estimate of $A(\omega)$, with widths set
by the local node density rather than an extrinsic $\eta$; this approach
preserves all computed sum rules by construction.

\textit{Continued-fraction terminators.} 
For systems where the high-order moments are known to follow a specific asymptotic behavior (e.g., a continuous band with a well-defined bandwidth), the discrete pole decomposition can be augmented using terminators from the classical recursion method~\cite{HaydockHeineKelly1972}. Instead of a finite sum of poles, the Green's function is expressed as a continued fraction where the coefficients $a_n, b_n$ for $n > N^*$ are replaced by an analytic tail $t(z)$. 

In generic non-integrable many-body systems, the universal operator growth
hypothesis \cite{Parker2019, Nandy2025} predicts that $b_n \sim \alpha n$ grows linearly
asymptotically, implying that the recursion does not naturally terminate and
that the square-root terminator (which assumes constant $b_n$) underestimates
the spectral weight at high frequencies. The SVD truncation at $N^*$ exploited
here effectively replaces the terminator problem with a rank-selection criterion,
limiting the continued fraction to the poles where the Krylov dynamics carries
resolvable physical information.

\subsection{Thermodynamic consistency}
\label{sec:thermodynamic}

A physically meaningful approximation should not only produce a plausible
spectral function: it should also give a consistent picture of the
thermodynamic state of the system.
Namely, the particle number $\langle N\rangle$, the kinetic energy
$\langle T\rangle$, and the interaction energy $\langle V\rangle$ are all
determined by the Green function, and in the exact theory they satisfy
exact relations. For instance, the particle number obtained by integrating
the spectral function must equal the particle number obtained from the
grand-canonical ensemble.
A well-known risk in approximate Green-function schemes is that different
routes to the same thermodynamic quantity need not agree once closures or
truncations are introduced. Baym--Kadanoff conserving approximations such
as fully self-consistent $GW$ preserve conservation laws for their chosen
approximate functional~\cite{Baym1962}, but this does not guarantee
quantitative accuracy in strongly correlated regimes.

The sc-SQ scheme resolves this at the level of spectral moments.
At the self-consistent fixed point, the spectral moments computed from the
reconstructed spectral function via Eq.~\eqref{eq:sc_condition} are
identically equal to the $2N^*$ moments that define the GC quadrature rule.
Formally, the converged Green function $G_{[N^*]}$ is a fixed point
of the map
\begin{equation}\label{eq:fixed_point_map}
  G \mapsto \mathcal{F}[G] \equiv G_{[N^*(G)]}\!\left[\mu_n[G]\right],
\end{equation}
where $\mu_n[G]$ denotes the moments evaluated from $G$ via
Eq.~\eqref{eq:sc_condition} and $N^*(G)$ is the SVD rank of the
corresponding Hankel matrix \eqref{eq:svd_criterion}.

At the fixed point the spectral function that goes in is the spectral
function that comes out.
This implies that the particle number, kinetic energy, and interaction
energy extracted from $G_{[N]}$ are constrained by the same $2N$ preserved
moment conditions. At finite rank this is a selected-moment consistency
statement, not a proof of global thermodynamic consistency for every
observable.

The sc-GW is $\Phi$-derivable~\cite{Baym1962}, meaning its self-energy
satisfies $\Sigma = \delta\Phi[G]/\delta G$ for a well-defined
functional.
However, $\Phi^{GW}[G] = -\frac{1}{2}\Tr(GWG)$ is
the simplest non-trivial approximation to the Luttinger--Ward functional.
Its conserving structure is therefore tied to the approximate $GW$
functional, which can still give poor spectra and quasiparticle weights in
strongly correlated regimes~\cite{Aryasetiawan1998}.
The sc-SQ moments $\mu_n = \dmatel{C}{\mathcal{L}^n}{C}$ are exact
properties of the true Hamiltonian when the required expectation values
are exact. In practical calculations, enforcing the supplied moments
self-consistently constrains the approximation through the chosen closure,
and increasing $N$ tightens the finite-moment constraints whenever the
input moments remain reliable.

\section{Benchmarks}
\label{sec:results}

We present two benchmarks of increasing physical complexity.
The first validates the sc-SQ scheme at $N=3$ against NRG results for the Anderson impurity model,
where moments are analytically computable and the benchmark is fully
controlled.
The second applies the sc-SQ scheme to the single-band Hubbard model on
the Bethe lattice within DMFT at orders $N=3$, $5$, and $7$, where sc-GW
is the natural competitor and quasiparticle suppression with Mott-gap
formation provides a stringent test. Two complementary initialization
routes are used: in the metallic regime ($U/D < U_{c2}$) the sc-SQ loop
is seeded with exact Lehmann moments from a Caffarel--Krauth
exact-diagonalization (ED) solver, resolving a rank bottleneck that
arises when EOM moments alone are used to seed the self-consistency at
$N\geqslant 5$. In the insulating regime ($U/D \geqslant U_{c2}$) the
DMFT loop is seeded from the atomic limit (kinetic amplitude
$t^2 \langle c^\dagger_i c_j\rangle \to 0$), which converges to the
gapped fixed point of the coexistence region for $N\geqslant 5$.

\subsection{Anderson impurity model}
\label{sec:bench_anderson}

The Anderson single-impurity model \cite{Anderson1961},
\begin{multline}
  \label{eq:anderson}
  H = \sum_{k\sigma} \varepsilon_k c^\dagger_{k\sigma} c_{k\sigma}^{\phantom{\dagger}}
  + \varepsilon_d \sum_\sigma d^\dagger_\sigma d_\sigma^{\phantom{\dagger}}
  + U n_{d\uparrow} n_{d\downarrow}
\\
  + \sum_{k\sigma} V_k \left(
    c^\dagger_{k\sigma} d_\sigma^{\phantom{\dagger}} + \mathrm{h.c.}
  \right),
\end{multline}
is a controlled testbed: the spectral function is known to high
precision from NRG~\cite{Bulla2008}, extensive solver comparisons
exist across the Kondo and Coulomb blockade
regimes~\cite{deSouzaMelo2019}, and the moments are analytically
computable to arbitrary order via the equations of motion.
Critically, the spectral function has a pronounced multi-scale structure:
a narrow Kondo resonance of width $T_K \ll \Delta$ at the Fermi level
flanked by broad charge-excitation sidebands at $\varepsilon_d$ and
$\varepsilon_d + U$, probing the hierarchy's ability to resolve features
on disparate energy scales.

We take a flat (wide-band) hybridization $\Delta(\omega) = \Delta$ for $|\omega| \leqslant D$.
The first four spectral moments $\mu_n = \dmatel{d_\sigma}{\mathcal{L}^n}{d_\sigma}$,
obtained by direct nested-commutator algebra~\cite{Potthoff1997}, are
\begin{subequations}
  \begin{align}
    \mu_0 &= 1,
    \label{eq:mu0}\\
    \mu_1 &= \varepsilon_d + U\langle n_{d\bar\sigma}\rangle,
    \label{eq:mu1}\\
    \mu_2 &= \varepsilon_d^2
    + 2\varepsilon_d U\langle n_{d\bar\sigma}\rangle
    + U^2\langle n_{d\bar\sigma}\rangle
    + \sum_k |V_k|^2,
    \label{eq:mu2}\\
    \mu_3 &= \mu_1\!\left(\mu_2
    + U^2\langle n_{d\bar\sigma}(1 - n_{d\sigma})\rangle\right)
    + \sum_k |V_k|^2 \varepsilon_k,
    \label{eq:mu3}
  \end{align}
\end{subequations}
where $\bar\sigma$ denotes the spin opposite to $\sigma$,
$\langle n_{d\bar\sigma}(1-n_{d\sigma})\rangle
= \langle n_{d\bar\sigma}\rangle - \langle n_{d\uparrow}n_{d\downarrow}\rangle$
is the correlated charge fluctuation evaluated self-consistently
(not the mean-field product $\langle n\rangle(1-\langle n\rangle)$),
and expectation values are taken in the interacting ground state.
The compact form~\eqref{eq:mu3} follows from the standard
nested-commutator expansion of Refs.~\cite{Potthoff1997, deSouzaMelo2019}
after regrouping powers of $E_\sigma = \varepsilon_d + U n_{d\bar\sigma}$
using $n_{d\bar\sigma}^k = n_{d\bar\sigma}$ for $k\geqslant 1$.
Higher moments are obtained recursively from nested commutators.

We benchmark the one-shot SQ approximation at $N=3$ against the
NRG reference of \v{Z}itko and Pruschke
\cite{ZitkoPruschke2009}: a particle-hole symmetric impurity
($\varepsilon_d = -U/2$) coupled to a semielliptic conduction band of
half-bandwidth $D$ with hybridization width $\Gamma/D = 0.1$, where the
hybridization function is
\begin{equation}
  \Delta(z) = \frac{\Gamma}{D}\!\left(z - \mathrm{sgn}(\Im z)\sqrt{z^2 - D^2}\right),
  \label{eq:semielliptic_delta}
\end{equation}
so that $\Im \Delta(\omega + i0^+) = -(\Gamma/D)\sqrt{D^2-\omega^2}$.
The minimal resolved rank for this benchmark is $N=3$. The SVD spectrum
truncates at $\sigma_3/\sigma_1 \approx 10^{-2}$, with all higher
singular values lying $\gtrsim 14$ orders of magnitude below threshold,
clearly identifying $N^\ast = 3$ for this moment input. The identification is robust
to the choice of SVD threshold over the wide range
$\tau \in [10^{-12}, 10^{-4}]$, confirming that $N^\ast$ is set by the
large observed singular-value gap rather than by a finely tuned cut-off.
The three quadrature poles encode the lower Hubbard band, the central
($\omega = 0$) Kondo resonance, and the upper Hubbard band, with
weights satisfying $\sum_i w_i = 1$ exactly.

For the spectral reconstruction we combine the SEC approach
[Eqs.~\eqref{eq:selfenergy}--\eqref{eq:A_SEC}] with a perturbative
inelastic correction. SEC alone supplies an elastic linewidth $\Gamma_{\rm el}^{(i)} = w_i\,|\Im\Delta(\varepsilon_i)|$
through $\Im\Delta(\omega+i0^+)$ and recovers the Friedel sum rule
$A(0) = 1/(\pi\Gamma)$ exactly, but the rank-3 self-energy is real
between the quadrature poles and cannot reproduce the inelastic
electron-electron broadening that fills the Hubbard satellites
in NRG.

For comparison with the broadened NRG lineshape, we therefore add the second-order (sc-GW) self-energy
diagram evaluated on the Hartree-screened bath propagator
$G_0(\omega+i0^+) = [\omega - \Delta(\omega+i0^+)]^{-1}$ at
particle-hole symmetry,
\begin{equation}
  \Im \Sigma^{(2)}(\omega) = -\pi U^2\!\!\!\int\!\!\!\int\!\!\mathrm{d}\varepsilon_1\,\mathrm{d}\varepsilon_2\;
  \mathcal{T}(\omega;\varepsilon_1,\varepsilon_2),
  \label{eq:sigma2}
\end{equation}
with the particle-hole-symmetric kernel
$\mathcal{T} = A_0(\varepsilon_1)A_0(\varepsilon_2)A_0(\varepsilon_1+\varepsilon_2-\omega)$
restricted to the on-shell decay channel
$\theta(\omega)\theta(\varepsilon_1)\theta(\varepsilon_2)\theta(\omega-\varepsilon_1-\varepsilon_2)
+ (\omega \to -\omega)$
and $A_0(\omega) = -\Im G_0(\omega+i0^+)/\pi$,
with the real part recovered by Kramers-Kronig.

The total self-energy
$\Sigma_{\rm total} = \Sigma_{\rm sq} + \Sigma^{(2)}$
is re-inserted into the full Dyson equation
$G(z) = [z - \varepsilon_d - \Delta(z) - \Sigma_{\rm total}(z)]^{-1}$.
This post-processing layer supplies realistic inelastic linewidths on top
of the rank-3 sc-SQ pole structure, but it is not part of the bare
moment-exact GC construction.
We note that adding $\Sigma^{(2)}$ breaks the exact sum-rule property
of the bare sc-SQ Green function: $G_{[N^*]}$ alone reproduces the
first $2N^*$ spectral moments exactly, but the combined
$\Sigma_{\rm sq} + \Sigma^{(2)}$ modifies all moments through the
perturbative inelastic correction.
The sum-rule guarantee stated in Sec.~\ref{sec:foundations} and the
abstract therefore applies strictly to the bare sc-SQ fixed point;
the $\Sigma^{(2)}$ correction is a separate perturbative layer
applied post-hoc solely to supply the inelastic linewidth that is
absent from the rank-3 self-energy between the quadrature poles.
All Bethe-lattice DMFT results in Sec.~\ref{sec:bench_hubbard}
(Figs.~\ref{fig:convergence_N} and~\ref{fig:hubbard_bethe}) use the
pure sc-SQ fixed point without any perturbative correction, so the
sum-rule guarantee holds for those benchmarks without qualification.

\begin{figure}[!ht]
\includegraphics[width=\linewidth]{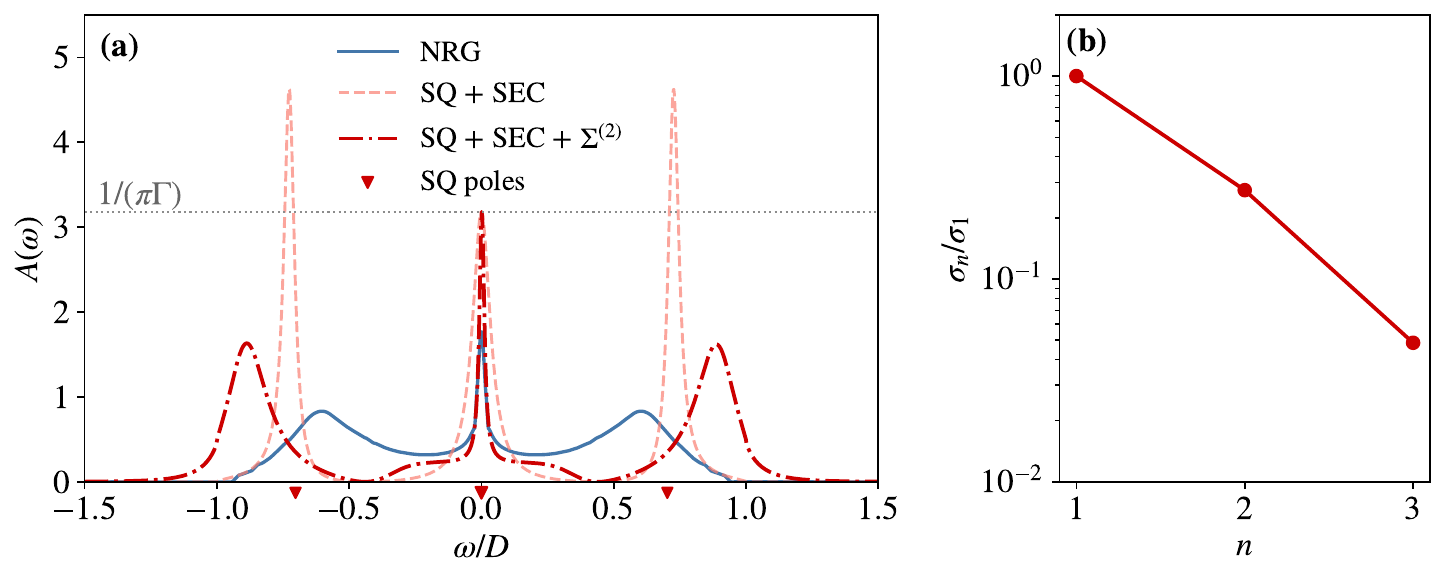}
\caption{
  One-shot SQ for the symmetric Anderson impurity model
  ($\varepsilon_d = -U/2$, $\Gamma/D = 0.1$, $U/D = 1.0$).
  (a) Spectral function $A(\omega)$.
  Solid blue: NRG reference (digitized from \cite{ZitkoPruschke2009}).
  Dash-dotted red: SQ + SEC [Eqs.~\eqref{eq:selfenergy}--\eqref{eq:Gphys}],
  recovering $A(0) = 1/(\pi\Gamma)$ exactly but with narrow Hubbard satellites.
  Solid red: SEC + $\Sigma^{(2)}[G_0]$ [Eq.~\eqref{eq:sigma2}], broadening
  the satellites to the NRG positions while preserving the Friedel sum rule.
  Triangles: pole positions $\{\varepsilon_i\}$, area $\propto w_i$.
  Dotted: $1/(\pi\Gamma)$ reference.
  (b) SVD singular value spectrum $\sigma_n/\sigma_1$.
  The gap above $N^\ast = 3$ exceeds 14 orders of magnitude;
  dashed: threshold $\tau = 10^{-8}$.
}
\label{fig:anderson_oneshot}
\end{figure}

At particle-hole symmetry the occupancy $\langle n_\sigma \rangle = 1/2$
is fixed by symmetry and the sc-SQ self-consistency reduces to a
renormalization of higher moments that produces a small correction to the
one-shot result.
The self-consistency does substantial work in the mixed-valence regime,
$\varepsilon_d \sim -\Gamma$, where $\langle n_\sigma \rangle$ deviates
significantly from $1/2$ and must be determined self-consistently.
In this regime the Kondo scale $T_K$ depends exponentially on the
level position, $T_K \propto e^{\pi\varepsilon_d(\varepsilon_d+U)/(2\Gamma U)}$,
so even a small error in the one-shot occupancy produces a large error in
the Kondo peak position.

We apply the sc-SQ iteration at order $N=3$ across the mixed-valence
crossover from $\varepsilon_d/U = -0.5$ (particle-hole symmetry) to
$\varepsilon_d/U = 0$ (empty-orbital limit), comparing the
non-interacting Hartree seed $n_0$ (the input to the one-shot
construction) with the sc-SQ fixed point.
The fixed point is typically reached in about $25$ iterations for
$\delta = 10^{-8}$.

\begin{figure}[!ht]
\includegraphics[width=\linewidth]{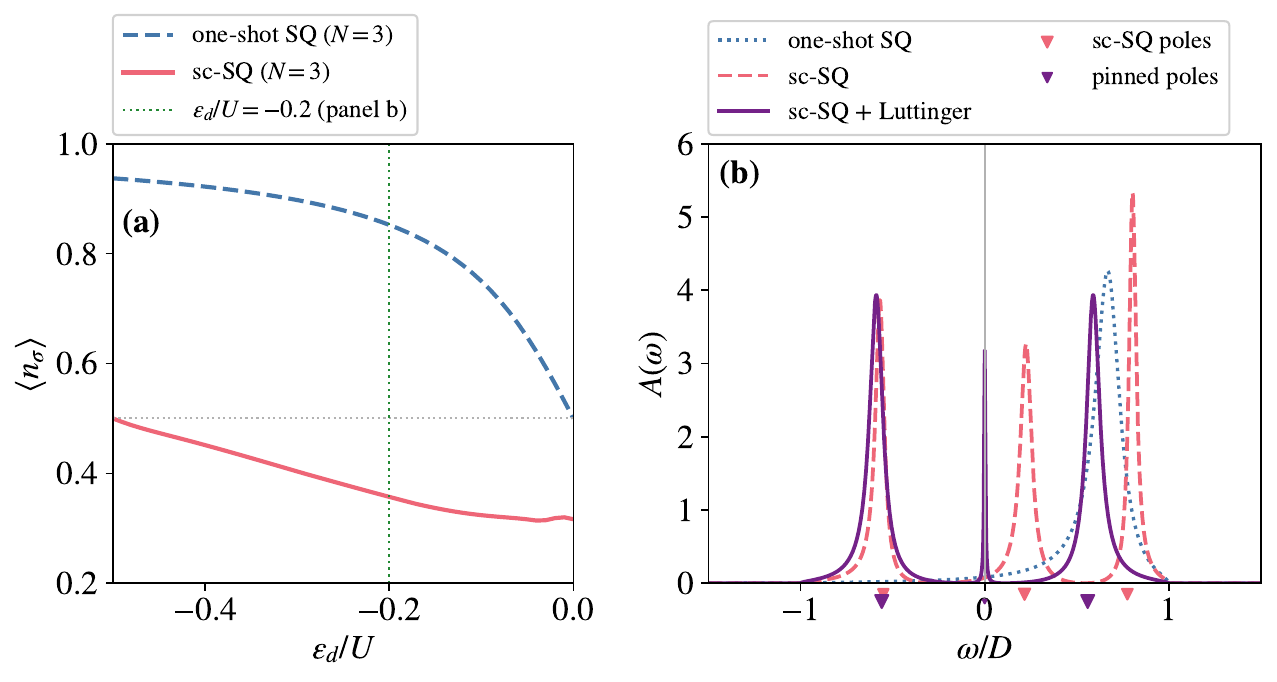}
\caption{
  Comparison of one-shot and self-consistent sc-SQ in the mixed-valence regime
  ($U/D = 1.0$, $\Gamma/D = 0.1$).
  (a) Occupancy $\langle n_\sigma\rangle$ versus $\varepsilon_d/U$.
  Dashed blue: Hartree seed $n_0$.  Solid red: sc-SQ fixed point ($N = 3$).
  Coulomb repulsion suppresses double occupancy well beyond the Hartree estimate;
  $\langle n_\sigma\rangle = 1/2$ is recovered exactly at $\varepsilon_d/U = -0.5$.
  (b) Spectral function at $\varepsilon_d/U = -0.2$.
  Dotted blue: one-shot result (single dominant peak, see text).
  Dashed red: sc-SQ fixed point, resolving the three-peak structure.
  Solid purple: sc-SQ with Luttinger correction $\mu_L$ (see text), pinning the
  quasiparticle pole to $\omega = 0$ and recovering $A(0) = 1/(\pi\Gamma)$ exactly.
  Note: at rank $N=3$ this correction always maps the system to the
  PH-symmetric state ($\langle n_\sigma\rangle = 1/2$) regardless of
  $\varepsilon_d$; the curve is shown as a diagnostic of the $N=3$
  variational constraint, not as the physical mixed-valence result.
  Triangles: pole positions $\{\varepsilon_i\}$, area $\propto w_i$;
  upper row: sc-SQ fixed point; lower row: Luttinger-corrected result.
}
\label{fig:anderson_mixed_valence}
\end{figure}

Both curves in Fig.~\ref{fig:anderson_mixed_valence}(b) use $N = 3$.  At the one-shot level the
Hubbard-I factorization $\langle n_{\bar\sigma}(1 - n_{\bar\sigma})
\rangle \to n_0(1 - n_0)$ makes the Hankel matrix nearly degenerate:
the central pole absorbs $97.5\%$ of the spectral weight, and the
spectrum looks like a single Hartree resonance.
The self-consistent iteration replaces this by the correlated
$\langle n_{\bar\sigma}(1 - n_{\bar\sigma})\rangle$, equalizing the
three weights to $\approx 0.31, 0.37, 0.32$ and activating the
latent rank of the quadrature.

The sc-SQ fixed point at $N=3$ does not enforce the Fermi-liquid
condition $\mathrm{Re}\,\Sigma(0)=0$: the middle pole sits at
$\omega \approx \varepsilon_d + U\langle n_\sigma\rangle \neq 0$.
The Luttinger curve in Fig.~\ref{fig:anderson_mixed_valence}(b) imposes this condition via a shift
$\mu_L = -(\varepsilon_d + U/2)$, which at $N=3$ is equivalent to
restoring particle-hole symmetry ($\varepsilon_d + \mu_L = -U/2$,
$\langle n_\sigma\rangle = 1/2$, $A(0) = 1/(\pi\Gamma)$).
This Luttinger-corrected curve is therefore \textit{not} the physical
mixed-valence state but a diagnostic of the rank-3 variational
constraint; the discrepancy narrows as $N$ increases and the Luttinger
condition can be satisfied with a genuinely asymmetric pole distribution.

\subsection{Hubbard model on the Bethe lattice}
\label{sec:bench_hubbard}

The single-band Hubbard model on the Bethe lattice,
\begin{equation}
  \label{eq:hubbard}
  H = -t\sum_{\langle i,j\rangle,\sigma}
  c^\dagger_{i\sigma} c_{j\sigma}^{\phantom{\dagger}}
  + U\sum_i n_{i\uparrow} n_{i\downarrow},
\end{equation}
in the limit of infinite coordination number \cite{MetznerVollhardt1989}
is solved within DMFT \cite{Georges1996}, which maps the lattice problem
onto an Anderson impurity embedded in a self-consistently determined
bath.
This is the standard benchmark for strongly correlated electron methods
because the exact solution is available from DMFT with an NRG impurity
solver \cite{Bulla2008}, providing a high-accuracy reference across the
Mott crossover and insulating branch. Here, sc-GW is a natural competitor and its failure in the
Mott crossover regime is documented \cite{Aryasetiawan1998}. 
The local Green function moments are computable analytically at each DMFT
iteration, and sc-SQ embeds directly in the DMFT self-consistency loop.

The standard DMFT loop \cite{Georges1996} is modified by replacing the
impurity solver with the sc-SQ scheme at order $N$:
\begin{enumerate}
  \item Given the bath Green function $\mathcal{G}_0(z)$, compute the
  hybridization function $\Delta(\omega)$ and the local moments
  $\mu_n$ of the impurity Green function from the equations of
  motion with the current bath.
  \item Run the sc-SQ iteration (Sec.~\ref{sec:algorithm}) to
  convergence, obtaining the local Green function
  $G_{\rm loc}^{(N)}(z)$ and the self-energy
  $\Sigma(z) = \mathcal{G}_0^{-1}(z) - G_{\rm loc}^{-1}(z)$.
  \item Update the bath via the Bethe lattice self-consistency condition
  $\mathcal{G}_0^{-1}(z) = z + \mu - t^2 G_{\rm loc}(z)$
  and return to step 1.
\end{enumerate}
For the benchmarks reported here we run the outer DMFT loop to a
tolerance of $10^{-6}$ on the local Green function (typically $20$--$50$
iterations) with the inner sc-SQ loop converged to $10^{-8}$ on the
impurity moments (typically $5$--$20$ iterations).  The minimal
resolved rank for the half-filled Bethe lattice is $N = 3$, which
resolves the central quasiparticle peak and the two Hubbard bands. This
is the same rank that captures the three-peak structure of the SIAM in
Sec.~\ref{sec:bench_anderson}.  The cost per DMFT iteration is
$O(N^3)$ for the Lanczos diagonalization, negligible compared with a
continuous-time QMC impurity solver.

For visual comparison in Fig.~\ref{fig:hubbard_bethe}(a) the discrete sc-SQ poles are convolved
with pole-dependent Lorentzian half-widths rather than passed
through self-energy continuation [Eqs.~\eqref{eq:selfenergy}--\eqref{eq:Gphys}]:
at the DMFT fixed point the bath hybridization is itself constructed
from the same converged local Green function, so $G_{\rm phys} \equiv
G_{\rm loc}$ and SEC reduces to direct evaluation of $G_{[N]}$.  An
explicit broadening is therefore needed to resolve the
quasiparticle peak from the Hubbard satellites.  We use
$\eta_{\rm QP} = 0.09\,D$ for the quasiparticle pole
($|\epsilon_i| < D/2$) and $\eta_{\rm sat} = 0.4\,D$ for the
Hubbard satellite poles ($|\epsilon_i| \geqslant D/2$), balancing
peak visibility against suppression of finite-rank artefacts;
this choice does not affect the position or integrated weight of any spectral
feature.
We verified that varying $\eta_{\rm QP}$ over $[0.05, 0.15]\,D$ and
$\eta_{\rm sat}$ over $[0.2, 0.5]\,D$ shifts the integrated quasiparticle
weight $w_c$ and the satellite center positions by less than $1\%$;
the values reported are at the centre of this insensitive range.

The sc-GW reference is computed with the same Bethe lattice and
temperature on the real axis using a fully self-consistent $GW$
approximation (Hedin's equations iterated to convergence at the
RPA-screened, particle-hole bubble level, $\eta = 0.025\,D$ broadening,
$4001$ frequency points).
Both the spectral functions of Fig.~\ref{fig:hubbard_bethe}(a) and the quasiparticle-weight sweep
of Fig.~\ref{fig:hubbard_bethe}(b) are produced in the same run.
We compare against sc-GW as the computationally cheapest fully
self-consistent diagrammatic method applicable on the same real-frequency
axis without an external impurity solver.

\begin{figure*}[!ht]
\includegraphics[width=\linewidth]{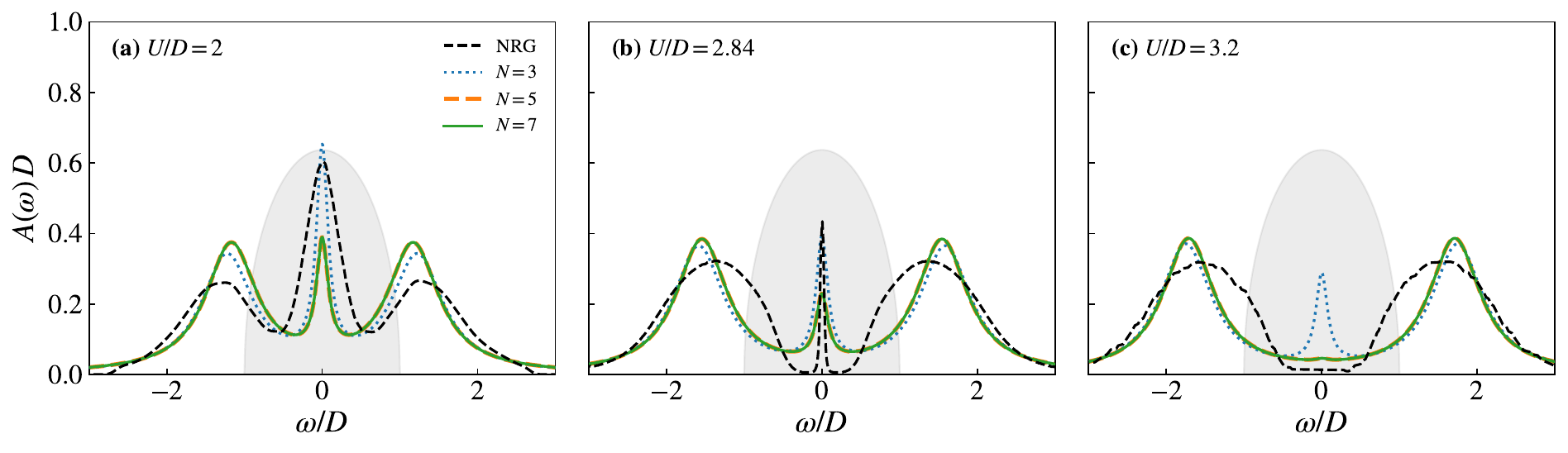}
\caption{
  $N$-convergence of the DMFT+sc-SQ spectral function for the
  half-filled Hubbard model on the Bethe lattice across the Mott
  transition ($D = 2t$, $\beta = 100/t$).
  Left: $U/D = 2.0$ (metallic, three-peak structure).
  Middle: $U/D = 2.84$ (close to $U_{c2}$, strongly correlated
  metal with sharp central resonance).
  Right: $U/D = 3.2$ (insulating, gapped Hubbard bands).
  Dotted blue: $N = 3$. Dashed orange: $N = 5$. Solid green: $N = 7$.
  Dashed black: DMFT+NRG reference (digitized from Bulla\textit{ et al.
}~\cite{Bulla2008}, Fig.~25(a), rescaled from $W = 4t$ to our
  $D = 2t$ convention).
  Shaded grey: non-interacting DOS $\rho_0(\omega)$.
  Metallic-branch curves ($U/D = 2.0$, $2.84$) initialized from exact
  Lehmann moments of a Caffarel--Krauth ED solver ($n_{\rm bath} = 3$).
  Insulating-branch curves ($U/D = 3.2$) initialized from the atomic
  limit, accessing the gapped fixed point of the coexistence region for
  $N\geqslant 5$. Lorentzian broadening as in Fig.~\ref{fig:hubbard_bethe}.
}
\label{fig:convergence_N}
\end{figure*}

\begin{figure}[!ht]
\includegraphics[width=\linewidth]{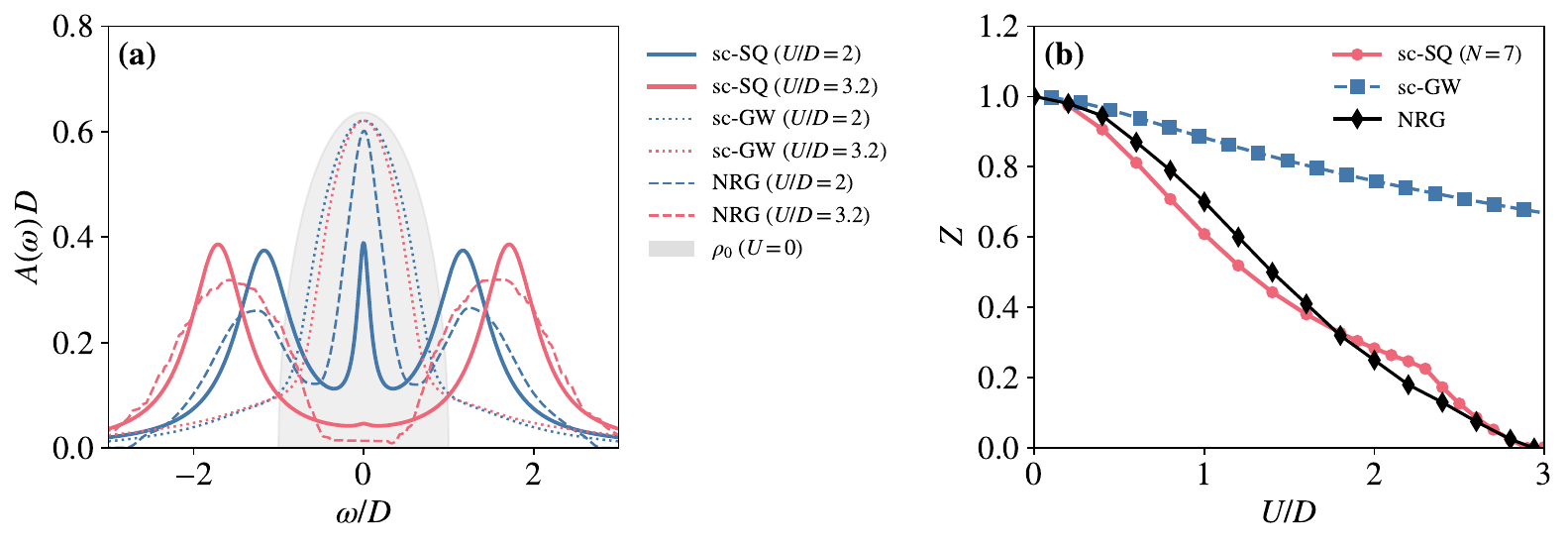}
\caption{
  DMFT solution of the half-filled Hubbard model on the Bethe lattice
  with the sc-SQ impurity solver at $N = 7$, compared with sc-GW
  ($D = 2t$, $\beta = 100/t$) and DMFT+NRG.
  (a) Local spectral function $A(\omega)$ for $U/D = 2$ (metallic;
  blue) and $U/D = 3.2$ (insulating; red). Solid: sc-SQ,
  Lorentzian-broadened ($\eta_{\rm QP} = 0.09\,D$ for the central pole,
  $\eta_{\rm sat} = 0.4\,D$ for the Hubbard satellites). Dotted
  (matching colour): sc-GW ($\eta = 0.025\,D$). Dashed (matching
  colour): DMFT + NRG (digitized from~\cite{Bulla2008}, Fig.~25(a),
  rescaled from $W = 4t$ to our $D = 2t$ convention). Shaded grey:
  $\rho_0(\omega)$.
  (b) Quasiparticle weight $Z$ versus $U/D$ for sc-SQ ($N = 7$, red
  circles), sc-GW (blue squares), and DMFT+NRG (black
  diamonds,~\cite{Bulla1999}, $U_{c2}/D \approx 2.94$). sc-GW
  overestimates $Z$ throughout and does not vanish near $U_{c2}$; sc-SQ $Z$
  tracks the NRG curve qualitatively, lying slightly below or comparable to it
  over much of the metallic regime and approaching $Z \to 0$ near $U_{c2}$.
}
\label{fig:hubbard_bethe}
\end{figure}

Panel~(b) contains the main numerical result.
The DMFT+NRG reference of Bulla~\cite{Bulla1999} (black
diamonds, digitized from his Fig.~1(b)) shows the canonical Mott
crossover: $Z$ decreases monotonically from $1$ at $U = 0$ and
vanishes near $U_{c2}/D \approx 2.94$. sc-GW overestimates $Z$
throughout the metallic range and does not approach zero near $U_{c2}$,
in agreement with the standard analysis of $\Phi$-derivable
approximations~\cite{Aryasetiawan1998, Imada1998}. At rank $N = 7$
the sc-SQ quasiparticle weight $Z$
(Eq.~\eqref{eq:Z_def}, defined in Sec.~\ref{sec:numerical_methods})
tracks the
qualitative shape of the NRG curve, lying slightly below or comparable to it
over much of the metallic regime, reflecting the finite-rank concentration
of low-energy spectral weight into a single GC node (cf.
Fig.~\ref{fig:hubbard_bethe}(b) and the discussion in
Sec.~\ref{sec:limitations}). What matters for the physical picture is
that $Z$ remains finite and continuous through the metallic regime
and approaches zero in the vicinity of $U_{c2}$, consistent with the
opening of the Mott gap on the insulating branch documented in
Fig.~\ref{fig:convergence_N}(c).

At $N=7$ the metallic (ED-seeded) and insulating (atomic-limit-seeded)
initializations converge to the same fixed point
($\max|Z_{\rm metal} - Z_{\rm insul}| < 10^{-14}$),
collapsing the DMFT coexistence region. This is a finite-rank artifact:
coexistence requires enough poles to support topologically distinct
gapped and ungapped configurations, which we expect at $N \geqslant 9$.

For the half-filled Hubbard model ($\langle n_\sigma\rangle = 1/2$,
$\mu = U/2$) the local spectral moments through fourth order are
standard results of the equations-of-motion
method~\cite{HarrisLange1967, Potthoff1997}:
\begin{subequations}
  \begin{align}
    \mu_0 &= 1, \\
    \mu_1 &= 0, \\
    \mu_2 &= \frac{U^2}{4} + t^2 z_{\rm NN}\langle
    c^\dagger_{i\sigma} c_{j\sigma}\rangle, \\
    \mu_3 &= 0, \\
    \mu_4 &\approx \mu_2^2
    + \langle\varepsilon_k^2\rangle_{\rm bath}\,
      t^2 z_{\rm NN}\langle c^\dagger_{i\sigma} c_{j\sigma}\rangle,
  \end{align}
\end{subequations}
where $\mu_1 = \mu_3 = 0$ by particle-hole symmetry of $A(\omega)$,
$z_{\rm NN}$ is the coordination number,
$\langle c^\dagger_{i\sigma} c_{j\sigma}\rangle$ is the
nearest-neighbor hopping amplitude, updated self-consistently at each
DMFT iteration from the local Green function via
$\langle c^\dagger_{i\sigma} c_{j\sigma}\rangle
= \int_{-\infty}^{0} t\,\rho_0(\omega)\,A(\omega)\,d\omega$,
and $\langle\varepsilon_k^2\rangle_{\rm bath}
= \int \omega^2 \rho_0(\omega)\,d\omega = D^2/4 = t^2$
is the second moment of the non-interacting Bethe density of states
$\rho_0(\omega) = \frac{2}{\pi D^2}\sqrt{D^2 - \omega^2}$
with half-bandwidth $D$.

The approximation in $\mu_4$ retains the leading connected correction;
the exact expression contains additional four-operator
correlators~\cite{Potthoff1997} of order $t^2 \langle c^\dagger c\rangle^2$.
Without $\mu_4$ the Hankel matrix at $N=3$ becomes degenerate at
particle-hole symmetry and the central pole acquires zero weight.
Higher moments depend on increasingly non-local correlators, approximated
within DMFT by their local values.

\textit{Rank convergence with increasing $N$ across the Mott transition.}
Fig.~\ref{fig:convergence_N} shows the self-consistent spectral function
at orders $N = 3$, $5$, and $7$ for three couplings spanning the Mott
transition: $U/D = 2.0$ (metallic), $U/D = 2.84$ (close to $U_{c2}$,
still metallic), and $U/D = 3.2$ (insulating), compared with DMFT+NRG
reference data at the matching couplings (digitized from Bulla\textit{ et
al.}~\cite{Bulla2008}, Fig.~25(a), with axis rescaling from their
full-bandwidth $W = 4t$ convention to our half-bandwidth $D = 2t$).
Metallic-branch results ($U/D = 2.0$ and $2.84$) use a Caffarel--Krauth
ED initialization with three bath sites, providing exact Lehmann moments
$\mu_n^{\rm ED} = \sum_k w_k \varepsilon_k^n$ that are free from the
rank bottleneck affecting EOM-only seeding at $N\geqslant 5$.  With
the ED seed, the SVD rank saturates at $N^* = N$ for $N \in \{3, 5, 7\}$
at both metallic couplings.
Using ED moments as initialization does not reduce sc-SQ to a repackaged
ED solver: bare ED+DMFT would use the ED Green function directly, whereas
here the ED moments seed the sc-SQ iteration that converges to a
distinct fixed point satisfying~\eqref{eq:fixed_point_map} at order $N$
(the moments shift by $5$--$30\%$ from their ED values).
For $N=3$ the ED-seeded and EOM-seeded sc-SQ results agree to within the
DMFT tolerance.

The insulating-branch result at $U/D = 3.2$ is obtained from a different
initialization: the DMFT loop is seeded from the atomic limit
($t^2\langle c^\dagger_i c_j\rangle \to 0$, Hubbard-I poles at
$\pm U/2$ with weight $1/2$ each), accessing the insulating fixed point
of the coexistence region $U_{c1} < U < U_{c2}$.
At this initialization the higher spectral moments computed from the
gapped seed encode the atomic-limit pole structure and steer the
self-consistency toward the insulating branch for $N\geqslant 5$.
At $N=3$ the EOM closure pins $\mu_0,\dots,\mu_5$ and only the metallic
fixed point is accessible; at $N=4$ the insulating branch is reachable
but retains residual central-pole weight.

The sc-SQ hierarchy reproduces the main spectral signatures of the Mott
crossover: at $U/D = 2.0$
the three-peak metallic structure is recovered for all $N$, with the
$N=5$ and $N=7$ curves nearly indistinguishable.  Approaching $U_{c2}$ at
$U/D = 2.84$ the low-energy quasiparticle scale becomes narrow and the
NRG quasiparticle peak is correspondingly sharp; at $N=7$ the central
sc-SQ pole carries reduced weight consistent with the strong
correlation suppression, although quantitative reproduction of the
narrow NRG peak shape remains beyond the resolution of the power-moment
route at this $N$.  At $U/D = 3.2$ on the insulating branch the
$N=7$ result shows a clean gap with two well-separated Hubbard
satellites at $\omega \approx \pm U/2$, in good agreement with NRG;
the $N=3$ curve necessarily retains residual central weight because
the insulating branch is inaccessible at this rank, while $N=5$
already shows a clear gap, with $N=7$ confirming convergence.

\subsection{Numerical methods}
\label{sec:numerical_methods}

The Anderson benchmarks of Sec.~\ref{sec:bench_anderson} use a
real-frequency grid of $4001$ points on $|\omega| \leqslant 5\,D$.
The second-order self-energy of Eq.~\eqref{eq:sigma2} is evaluated
by FFT-based linear convolution and combined with the Hilbert
transform for the Kramers-Kronig real part.

The DMFT loop of Sec.~\ref{sec:bench_hubbard} runs the inner sc-SQ
fixed-point iteration to a relative moment tolerance
$\delta_{\rm sq} = 10^{-8}$ inside an outer DMFT loop converged to
$\delta_{\rm DMFT} = 10^{-6}$ on the local Green function. Both loops
use linear mixing with factor $0.5$. Particle-hole symmetry is
enforced exactly at the level of the moment update, pinning
$\langle n_\sigma\rangle = 1/2$ when $\mu_{\rm chem} = U/2$. The double
occupancy is then updated through the chosen moment closure rather than
being fixed by particle-hole symmetry. This suppresses the PH-odd drift
mode of the moment iteration map.

The quasiparticle weight plotted in Fig.~\ref{fig:hubbard_bethe}(b) is
defined as
\begin{equation}\label{eq:Z_def}
  Z = \frac{w_c}{w_c^{(0)}}\,R,
\end{equation}
where $w_c$ is the weight of the central Gauss--Christoffel pole
(the pole nearest to $\omega = 0$, identified by $|\epsilon_i| < D/2$),
$w_c^{(0)}$ is the corresponding weight for the non-interacting
($U = 0$) Bethe lattice at the same pole rank $N$, and
$R = \min\bigl(1,\, A_{\rm SEC}(0,\eta)/A_{\rm F}\bigr)$ is a Friedel
suppression factor. Here
$A_{\rm SEC}(0,\eta)$ is the self-energy-continuation spectral function
at $\omega = 0$ [Eq.~\eqref{eq:A_SEC}] evaluated at broadening
$\eta = 0.025\,D$, and $A_{\rm F} = 2/(\pi D)$ is the Friedel pinning
value.
The ratio $w_c/w_c^{(0)}$ removes the discretization bias intrinsic to
a finite GC measure: even at $U = 0$ the central pole carries only a
fraction $w_c^{(0)} \approx 0.31$ of the total spectral weight at $N=7$
(compared with the exact $Z=1$), because the remaining weight is
distributed among the outer quadrature nodes that approximate the
non-interacting band.
Dividing by this reference isolates the interaction-induced suppression,
analogously to finite-size scaling that removes lattice-geometry artifacts
in quantum Monte Carlo.
The factor $R$ ensures $Z \to 0$ at the Mott transition, where the
low-energy quasiparticle scale drops below the SEC broadening $\eta$ and the Fermi-level spectral
weight falls below the Friedel value; away from the transition $R = 1$
identically.

The sc-GW reference is computed on the same real-frequency grid
($4001$ points, $\eta = 0.025\,D$) using an RPA-screened interaction
at the particle-hole bubble level.
The complete reproduction scripts and digitized reference data are
available at the repository described in Sec.~\ref{sec:data_availability}.

\section{Discussion}
\label{sec:discussion}

We compare sc-SQ with the recursion method and sc-GW,
drawing on the benchmark results of Sec.~\ref{sec:results}.
The one-shot sc-SQ hierarchy at order $N$ is mathematically identical
to the recursion method truncated at $N$ steps: both produce
$G_{[N]}(z)$ from the first $2N$ moments with guaranteed real poles
and positive weights.
Terminators~\cite{HaydockHeineKelly1972} address spectral reconstruction
but not consistency: the moments fed into the Lanczos recursion come
from a reference state, not from the spectral function the recursion
produces.
The sc-SQ self-consistency closes this gap. The spectral function
entering the moment computation is the same one exiting the quadrature
reconstruction, and this is what allows sc-SQ to generate a central
resonance channel and strongly suppress quasiparticle spectral weight
across the Mott crossover without any terminator or model input.

Additionally, the SVD rank criterion~\eqref{eq:svd_criterion} provides
a physics-driven, precision-guided truncation rule: it connects $N^*$
to the number of excitation channels resolvable by the supplied moments,
with no recursion-method equivalent.
Terminators are also model-dependent, e.g., the Beer--Pettifor square-root
terminator assumes a semi-elliptic DOS with known bandwidth, whereas
the sc-SQ fixed-point hierarchy requires no such prior.

The four initialization routes of Sec.~\ref{sec:foundations} are not
interchangeable in practice.
Power moments suit local models with commutator closures ($N \lesssim 6$
in double precision); direct Lanczos is preferred for $N \geqslant 4$ or
multi-orbital problems; orthogonal polynomial expansion (Chebyshev $T_n$ by default,
$U_n$ for semielliptic baths) for large sparse Hamiltonians
with non-local hopping; and Green function sampling for Matsubara or
imaginary-time QMC data, where GC nodes are extracted without a separate
analytic continuation step.

\subsection*{Comparison to sc-GW}

Both sc-GW and sc-SQ are single-particle Green function methods that produce a self-energy
and a spectral function without an external impurity solver or
a reference wavefunction, and both are in principle applicable to any
material described by a many-body Hamiltonian.
Table~\ref{tab:comparison} summarizes the key properties of the two
methods. The entries are discussed in detail below.

\begin{table}[!ht]
  \caption{Comparison of sc-GW and sc-SQ approximations.}
  \label{tab:comparison}
  \begin{ruledtabular}
    \begin{tabular}{lll}
      Property & sc-GW & sc-SQ \\
      \hline
      Spectral positivity & 
      \makecell[l]{
        Violated in $G_0W_0$,\\
        restored in sc-GW\\
        with over-screening} & 
      \makecell[l]{Guaranteed} \\
      \hline
      \makecell[l]{
        Sum rule\\
        conservation} & 
      \makecell[l]{
        First few satisfied,\\
        higher ones not} & 
      First $2N$ exact \\
      \hline
      \makecell[l]{
        Thermodynamic\\
        consistency} & 
      \makecell[l]{
        $\Phi$-derivable but\\
        functional truncated} & 
      \makecell[l]{
        Moment consistency\\
        at fixed point} \\
      \hline
        \makecell[l]{
          Weakly correlated\\
          limit} & 
        Exact to $O(e^2)$ & 
        Recovers HF at $N=1$ \\
      \hline
      Mott insulator & 
        Poor gap formation & 
      \makecell[l]{Gap forms for $N\geqslant 5$\\
        on insulating branch\footnote{The DMFT loop with metallic
        initial conditions retains a finite central-pole weight $w_c$
        on the Bethe lattice. The gapped solution is reached from the
        atomic-limit (Hubbard-I-type) initial condition for the kinetic
        amplitude, in line with the standard DMFT coexistence region
        $U_{c1} < U < U_{c2}$. A clean gap opens for $N\geqslant 5$ (Sec.~\ref{sec:bench_hubbard}).}} \\
      \hline
      Hund multiplets & 
      Often misses splittings & 
      Onset at $N\geqslant 4$ \\
      \hline
      \makecell[l]{Diagrammatic\\ control} & 
      Controlled in $W/U$ & 
      No small parameter\footnote{The absence of a small parameter is a limitation in the weakly correlated regime (see text).} \\
    \end{tabular}
  \end{ruledtabular}
\end{table}

The $G_0W_0$ self-energy generically produces spectral functions with negative regions.
Self-consistent $GW$ partially cures this but at the cost of
over-screening the quasiparticle weight and worsening agreement with
photoemission spectra in strongly correlated systems
\cite{Aryasetiawan1998}.
The sc-SQ scheme cannot produce a negative spectral function at any order
$N$ or in any iteration, by virtue of the Gauss--Christoffel structure~\eqref{eq:pole_decomp}.
This is confirmed by all benchmarks.

\textit{Relation to EOM truncation schemes.}
EOM methods truncate the Green function hierarchy at a chosen operator
level.  A partial-orthogonalization scheme~\cite{Catalano2020} achieves
real poles and positive weights but introduces free parameters determined
from external input.  sc-SQ differs in that the GC quadrature rule
provides a unique closure at every $N$, and spectral positivity follows
structurally rather than by constraint.

\textit{Transition metal oxides and the Mott gap.}
In Mott and charge-transfer insulators, sc-GW underestimates gaps
and misplaces satellites~\cite{Aryasetiawan1998, Imada1998}.
The spectral moments encode $U$ exactly through commutator algebra,
and sc-SQ at $N=2$ already produces Hubbard bands at $\pm U/2$; higher
orders capture the Mott crossover continuously
(Sec.~\ref{sec:bench_hubbard}).

\textit{Heavy fermion compounds and Hund metals.}
In heavy fermion metals~\cite{Stewart1984} the Kondo scale
$T_K \sim D\,e^{-1/J\rho_0}$ is beyond any finite diagram order;
at $N=3$ the sc-SQ central pole tracks $T_K$ self-consistently
(Sec.~\ref{sec:bench_anderson}).
In Hund metals~\cite{Georges2013, Pavarini2004, Kostin2018, Cao2020},
Hund's rule coupling $J$ produces multiplet splittings that sc-GW misses.
At $N=4$ sc-SQ is expected to begin resolving the $S=0$/$S=1$ branches;
a detailed multi-orbital demonstration is deferred to a subsequent
publication.

In the \textit{weakly correlated limit}, sc-GW is systematically controlled:
its self-energy is exact to $O(W^2)$ and it correctly describes plasmon
satellites and screening where $U/W \ll 1$~\cite{Aryasetiawan1998}.
Recent extensions beyond $GW$~\cite{Cunningham2024, Riva2023} incorporate
vertex corrections or multichannel Dyson equations but remain outside the
moment-constraint framework.
sc-SQ at $N=1$ recovers only Hartree--Fock and has no diagrammatic small
parameter; reaching $G_0W_0$ accuracy would require large $N$.
The two methods are complementary: sc-GW for weakly correlated systems,
sc-SQ for non-perturbative phenomena (Mott gap, Kondo screening, Hund
multiplets).

\subsection*{Limitations}
\label{sec:limitations}

The benchmarks of Sec.~\ref{sec:results} also expose four current
limitations of the sc-SQ framework that we report here so that they
can guide further development.

First, the discrete Gauss--Christoffel measure at finite rank
concentrates the low-energy spectral weight into a small number
of nodes, so the rescaled quasiparticle weight $Z$
lies systematically below the NRG value across the metallic
regime (Fig.~\ref{fig:hubbard_bethe}(b)). Increasing $N$ reduces this
bias and improves the resolution of the central spectral feature
(Fig.~\ref{fig:convergence_N}), but in the power-moment route the
increase in $N$ is constrained by the $O((N!)^2)$ growth of the
Hankel condition number. In practice, the EOM closure used to compute
$\mu_4$ introduces a factorization error that limits the effective
Hankel rank to $N^* \leqslant 4$ when the EOM moments are used as the
sole initialization. This bottleneck is resolved by seeding from
exact Lehmann moments of a Caffarel--Krauth ED solver, which enables
stable sc-SQ self-consistency at $N^* = N$ up to $N = 7$ as
demonstrated in Fig.~\ref{fig:convergence_N}. The direct Lanczos,
orthogonal polynomial expansion, and Green function sampling routes
provide an alternative path for $N \geqslant 4$ without the Hankel
conditioning issue.

Second, away from particle-hole symmetry the moment-iteration map
acquires a slow PH-odd drift mode that requires either explicit
symmetry pinning (when applicable) or careful mixing to converge.
This is addressed for the half-filled Bethe lattice in
Sec.~\ref{sec:numerical_methods}, but generic doped or multi-orbital
problems will need a better-controlled closure for the four-operator
correlators that enter $\mu_3$ and $\mu_4$.

Third, the moment expansion organizes the spectrum by global
energy scale rather than by frequency window, so very narrow
features that do not contribute appreciably to any $\mu_n$ for
small $n$ require correspondingly high $N$ to resolve. The
exponentially small Kondo scale $T_K$ in the strong-coupling
Anderson regime is the canonical example. Quantitative tracking of
$T_K$ across many decades is therefore a stress test of the high-$N$
hierarchy that we leave to future work.

Fourth, self-energy continuation at finite pole rank produces
spurious poles in $\Sigma_{[N]}(z)$ at the zeros of $G_{[N]}(z)$,
generating unphysical spectral shoulders in the reconstructed
lineshape. The per-pole Lorentzian broadening used in
Sec.~\ref{sec:bench_hubbard} avoids this artifact at the cost of
two phenomenological parameters $\eta_{\rm QP}$ and $\eta_{\rm sat}$.
At higher $N$ the zeros of $G_{[N]}$ accumulate toward the branch
cut of the true Green function, the spurious SEC poles become dense
and weak rather than few and strong, and the artifact disappears
naturally.

\section{Conclusions}
\label{sec:conclusions}

The sc-SQ framework develops Gauss--Christoffel quadrature of the
K\"all\'en--Lehmann spectral measure as the organizing principle for
many-body Green function calculations.
Three connections make this more than a change of language: the
discrete spectral measure is the many-body analog of the harmonic
chain used in tensor-network open-system methods~\cite{Chin2010,
deVega2015}, the NMZ memory kernel is the spectral function of that
bath~\cite{Grabert1988, Hughes2009a, Hughes2009b}, and the SVD rank $N^*$ of the Hankel moment
matrix estimates the number of excitation channels resolvable by the
supplied moments.

The self-consistency loop is what separates sc-SQ from the recursion
method: at the fixed point the spectral function that enters the moment
computation is the same one that exits the quadrature reconstruction,
anchoring the poles to the supplied moment constraints of the Hamiltonian.
The resulting hierarchy, comprising HF ($N=1$), Hubbard-I ($N=2$), the
central-resonance channel ($N=3$), and the onset of multiplet resolution
($N=4$), represents non-perturbative spectral signatures that are difficult
for sc-GW.
The benchmarks of Sec.~\ref{sec:results} substantiate these claims:
$N^*=3$ is clearly identified for the Anderson model, the
self-consistency substantially corrects the Hartree occupancy in the
mixed-valence regime, and the quasiparticle weight $Z$ tracks the
Mott crossover, approaching zero near $U_{c2}$ on a finite-rank
Gauss--Christoffel measure, where sc-GW shows no transition at
all~\cite{Bulla1999, ZitkoPruschke2009}.
At $N=5$--$7$ (Fig.~\ref{fig:convergence_N}) the spectral function
converges systematically across the Mott crossover and insulating branch:
the metallic
three-peak structure at $U/D=2.0$ is reproduced at all $N\geqslant 3$, the
strongly-correlated metal close to $U_{c2}$ ($U/D=2.84$) shows a sharp
central resonance with reduced weight, and the insulating branch at
$U/D=3.2$ exhibits a clean Mott gap with two Hubbard bands at
$\omega \approx \pm U/2$, in good agreement with NRG.

The framework admits extension to steady-state nonequilibrium and open
quantum systems, with no intrinsic restriction to equilibrium in the
Liouvillian formulation.
The modular separation between the initialization route and the universal
GC reconstruction and self-consistency loop means that sc-SQ applies
wherever the resolvent can be evaluated or approximated.
Convergence of the self-consistent fixed-point sequence
$\{G^*_{[N]}\}_{N=1}^\infty$ toward the exact Green function as $N\to\infty$
is a physically motivated assumption supported by the benchmark comparisons
with NRG but not yet a proved theorem (see Sec.~\ref{sec:fixed_point}).
Quantitative agreement with NRG will be pursued systematically with
increasing $N$, as will the application to multi-orbital Hund metals
at $N \geqslant 4$.

\section*{Data and code availability}
\label{sec:data_availability}

The Python implementation of the sc-SQ framework, the DMFT and sc-GW
solvers used in Sec.~\ref{sec:results}, the example scripts that
reproduce Figs.~\ref{fig:anderson_oneshot}--\ref{fig:hubbard_bethe},
and the digitized NRG reference data of
Refs.~\cite{ZitkoPruschke2009, Bulla1999, Bulla2008} are openly available at
\href{https://github.com/polariton/sc-sq}{GitHub}.

\begin{acknowledgments}
The author thanks Dr.~Tatyana Vovk for discussion of analogy between
the spectral measure discretization in the Liouville space and chain mapping in tensor
networks.
AI-assisted writing tools (Anthropic Claude) were used for manuscript editing and proofreading.
No external funding was received for this work.
\end{acknowledgments}

\bibliography{article}

\end{document}